\documentclass[twocolumn,
              superscriptaddress,
              floatfix,
              longbibliography, 
              pra]{revtex4-2}  

\usepackage{graphicx}
\usepackage{dcolumn}
\usepackage{bm}
\usepackage{physics}
\usepackage{dsfont}
\usepackage{mathtools}
\usepackage{algorithm}
\usepackage{algpseudocode}
\usepackage[caption=false]{subfig}
\usepackage{color}
\usepackage{comment}
\usepackage{hyperref}
\usepackage{longtable} 
\usepackage{appendix}
\usepackage{url}
\usepackage{xcolor}
\usepackage{amsmath} 
\usepackage{amssymb} 
\usepackage{longtable} 
\usepackage{float}
\usepackage{xparse}
\usepackage[normalem]{ulem}
\usepackage{xr}
\definecolor{darkred}{rgb}{0.55, 0.0, 0.0}
\definecolor{darkgreen}{rgb}{0.0, 0.55, 0.0}
\definecolor{darkblue}{rgb}{0.0, 0.0, 0.55}
\hypersetup{colorlinks=true,urlcolor=darkblue,linkcolor=darkred,citecolor=darkgreen}
\makeatletter

\newtheorem{theorem}{Theorem}
\DeclareMathOperator*{\argmax}{arg\,max}

\DeclareMathOperator*{\sups}{sup\, }

\DeclarePairedDelimiter\ceil{\lceil}{\rceil}

\def\RIM{\text{RIM}}
\def\ARIM{\text{ARIM}}

\begin{document}

\title{Statistically Characterizing Robustness and Fidelity of Quantum Controls and Quantum Control Algorithms}

\author{Irtaza Khalid}
\email{khalidmi@cardiff.ac.uk}
\affiliation{School of Computer Science and Informatics, Cardiff University, Cardiff, CF24 4AG, UK}

\author{Carrie A.\ Weidner}
\email{c.weidner@bristol.ac.uk}
\affiliation{Quantum Engineering Technology Laboratories, H.\ H.\ Wills Physics Laboratory and Department of Electrical and Electronic Engineering, University of Bristol, Bristol, BS8 1FD, UK}

\author{Edmond A.\ Jonckheere}
\email{jonckhee@usc.edu}
\affiliation{Department of Electrical and Computer Engineering, University of Southern California, Los Angeles, CA 90089, US}

\author{Sophie G.\ Shermer}
\email{lw1660@gmail.com}
\affiliation{Department of Physics, Swansea University, Swansea, SA2 8PP, UK}

\author{Frank C.\ Langbein}
\email{frank@langbein.org}
\affiliation{School of Computer Science and Informatics, Cardiff University, Cardiff, CF24 4AG, UK}

\date{\today}

\begin{abstract}
Robustness of quantum operations or controls is important to build reliable quantum devices. The \emph{robustness-infidelity measure} ($\RIM_p$) is introduced to statistically quantify in a single measure the robustness and fidelity of a controller as the $p$-th order Wasserstein distance between the fidelity distribution of the controller under any uncertainty and an ideal fidelity distribution. The $\RIM_p$ is the $p$-th root of the $p$-th raw moment of the infidelity distribution. Using a metrization argument, we justify why $\RIM_1$ (the average infidelity) is a good practical robustness measure. Based on the $\RIM_p$, an algorithmic robustness-infidelity measure (\ARIM) is developed to quantify the expected robustness and fidelity of controllers found by a control algorithm. The utility of the \RIM\ and \ARIM\ is demonstrated on energy landscape controllers of spin-$\tfrac{1}{2}$ networks subject to Hamiltonian uncertainty. The robustness and fidelity of individual controllers as well as the expected robustness and fidelity of controllers found by different popular quantum control algorithms are characterized. For algorithm comparisons, stochastic and non-stochastic optimization objectives are considered. Although high fidelity and robustness are often conflicting objectives, some high-fidelity, robust controllers can usually be found, irrespective of the choice of the quantum control algorithm. However, for noisy or stochastic optimization objectives, adaptive sequential decision-making approaches, such as reinforcement learning, have a cost advantage compared to standard control algorithms and, in contrast, the high infidelities obtained are more consistent with high \RIM\ values for low noise levels.
\end{abstract}


\maketitle

\section{Introduction}\label{sec:intro}

Fault-tolerance is crucial for quantum technology and presents a particular challenge for Noisy Intermediate-Scale Quantum (NISQ) devices~\cite{nisq}. Broadly, there are three proposed ways to deal with noise and errors and achieve fault-tolerance:
(1) via error correction protocols, e.g., Shor codes~\cite{shor1, shor2, shor3}, or syndrome measurements~\cite{shor4};
(2) using error mitigation schemes, e.g., reversing noisy dynamics~\cite{rev1, kemmerev1, otherrev2, errormit1}, active variational noise minimization~\cite{rev2}, or parametric modelling of architecture defects in trapped qubits~\cite{rev3, rev3phdth};
(3) robust solutions engineering, e.g., landscape shaping of the quantum control optimization problem in search of noise-free regions~\cite{noisefree1, noisefree2, holopara}, decoherence-free subspaces~\cite{decfreesubspaces, DSM_CDC}, or noise spectral density based filter functions~\cite{biercuk1, biercuk2}.
Uncertainties that require fault-tolerance in quantum devices have two flavors:
(a) interactions with the environment that lead to non-unitary dynamics;
(b) inaccuracies in the control model representing a specific physical implementation that affect the evolution but do not cause non-unitary evolution.

Standard quantum control methods for steering quantum devices mostly focus on finding controls that have high fidelity using mathematical models~\cite{qopt2, qopt3, blumelanalyticrobustness}. However, if the operation of quantum devices is subject to noise, high fidelity itself is insufficient to gauge performance of a control scheme, and extra effort is required to systematically search for solutions that are both, robust against noise and have high fidelity~\cite{statisticalrobustness1, statisticalrobustness2}. This requires a notion of robustness and ideally a single measure that can capture robustness and fidelity, enabling the identification and construction of more efficient methods to find controls that satisfy both properties.

In this paper, we introduce a general statistical diagnostic based on the Wasserstein distance of order $p$~\cite{villani} to evaluate the robustness and fidelity of quantum control solutions and the algorithms used to find them. This is applicable to any quantum control problem where the fidelity is a random variable with a probability distribution over $[0,1]$. The Wasserstein distance between probability distributions is a measure of the minimal costs of probability mass transport between two distributions. In Sec.~\ref{sec:method}, the $p$-th order \emph{Robustness-Infidelity Measure} ($\RIM_p$) is defined to quantify the robustness and fidelity of a quantum controller. It is the $p$-th order Wasserstein distance between the probability distribution for the fidelity induced by noise and the ideal distribution for a perfectly robust controller, described by a Dirac delta function at fidelity $1$. We show that the $\RIM_p$ is the $p$-th root of the $p$-th raw moment of the infidelity distribution -- a non-parametric measure independent of any particular assumption for the distribution.

Using measure-theoretic norm scaling relations between \RIM{}s of different order, we argue that $\RIM_1$, the average infidelity, is sufficient as a practical measure of robustness and fidelity. As such, it meshes with the average infidelity, already used in robust~\cite{ensemblebloch, robustprecisecontrol} and stochastic/adaptive quantum control settings~\cite{stochlearningcontrolensemble, learningcontrolensemble}. More generally, the $\RIM_p$ is related to the risk-tunable fidelity measure using a utility function, as introduced in Ref.~\cite{risksensitivecontrol}.

The \RIM\ has practical utility by allowing us to choose among similar, high-fidelity controllers, acting as a post-selector for robust controllers, agnostic of the algorithm used to find them. This may be computationally more efficient than optimizing the \RIM\ directly, as we see in Sec.~\ref{sec:uncons_algocomparison}. Moreover, it can also be adapted to compare the performance of control algorithms in finding not only high-fidelity but also robust controllers. To that end, we introduce an Algorithmic \RIM\ (\ARIM), averaging \RIM{}s over multiple controllers, in Sec.~\ref{sec:method}.

In Sec.~\ref{sec:numexp} we illustrate the \RIM\ and draw useful insights for robust quantum control by generating controllers for static energy landscape control of the XX Heisenberg model, exploiting the degree of freedom afforded by the existence of multiple optima in quantum control~\cite{multipleoptima}. We analyze their robustness properties and the performance of algorithms in finding effective controllers using four optimization algorithms representing different, commonly employed approaches:
(1) L-BFGS: a second-order gradient-based optimization using an ordinary differential equation model of the quantum system to compute the fidelity under perfect conditions~\cite{lbfgs};
(2) Proximal Policy Optimization (PPO): a model-free reinforcement learning algorithm, having no prior knowledge of the system~\cite{ppo};
(3) Nelder-Mead: a derivative-free simplex-based heuristic search method~\cite{neldermead};
and (4) Stable Noisy Optimization by Branch and Fit (SNOBFit): another derivative-free method that performs model-free learning by using regression to estimate gradients via a branch and fit method~\cite{snob}.
Here, (1) serves as a baseline for optimization over a noise-free fidelity objective functional under ideal conditions. (2) represents a machine learning approach with minimal knowledge. (3) and (4) are derivative-free methods to handle stochastic objective functionals. A detailed motivation for the choice of these algorithms is presented in Sec.~\ref{sec:algomotivation}. These choices are not exhaustive but serve as a diverse set of algorithms to which we apply the \RIM\ and \ARIM, illustrating their utility and giving some indication of the performance of common control algorithms for the specific robust control problem. The algorithms were implemented in Python. Specifically, we used the SciPy library for (1) and (3)~\cite{scipy}, and (4) is obtained from Ref.~\cite{scikitquant}. Our code and data are available at Ref.~\cite{robchar}.

Our experimental motivation is four-fold:
(A) by comparing the robustness of controllers without regard to the optimization algorithm, we wish to answer whether high fidelity implies high robustness using the \RIM\ of the individual controllers (Sec.~\ref{sec:individualcont}).
(B) by conducting a distributional comparison of controllers we wish to understand how likely it is that a given algorithm produces controllers in an ideal (no-noise) setting that are robust in noisy conditions (Sec.~\ref{sec:consistencystat}).
(C) to study the effect of training noise of the same nature as the robustness noise model applied \emph{during} optimization on an algorithm's ability to find robust controllers using the \ARIM\ (Sec.~\ref{sec:algocomparison}).
For a fair comparison, we conduct (B) and (C) with a fixed number of objective function calls allotted to each algorithm.
(D) In Sec.~\ref{sec:uncons_algocomparison}, we try to understand an algorithm's asymptotic ability to find robust controllers using the \ARIM\ through optimizing the \RIM\ by allowing unlimited objective function calls. We consider two settings in this scenario: stochastic and non-stochastic fidelity optimization. In the latter case we optimize over a fixed set of Hamiltonians sampled once according to a noise model, while in the former case the Hamiltonians are stochastically chosen at each objective function evaluation using the same noise model.

Our main numerical findings are:
\begin{itemize}
\item High fidelity controllers are not always robust, but the non-robust controllers can be filtered out using the \RIM.
\item Using a consistency statistic, we show that PPO controller infidelities (\RIM\ at no noise) are more correlated with \RIM\ values at low noise levels compared to the other algorithms. More generally, a strong signal in the consistency statistic predicts \RIM\ robustness while avoiding its explicit evaluation.
\item For constrained objective function calls, there appear to be problem-dependent optimal levels of noise that produce more robust controllers for PPO in contrast to L-BFGS, SNOBFit and Nelder-Mead.
\item Robust controllers with respect to certain noise models in the optimization objective are obtained by all algorithms for the non-stochastic optimization objective when there are no constraints on resources.
\item However, if the optimization objective is stochastic, the \ARIM\ improves asymptotically only for PPO. In either case, PPO requires fewer function calls compared to the other algorithms, which highlights the potential of adaptive sequential decision-making strategies like reinforcement learning for NISQ optimization problems, where not all uncertainty can be captured by non-stochastic objective functionals (e.g., shot noise).
\end{itemize}

Lastly, from the perspective of classical control~\cite{Safonov_Laub_Hartmann}, it is well known that accuracy conflicts with robustness through the $S+T=I$ formula, where $S$ is related to tracking error and $T$ to sensitivity of the tracking error to uncertainties. This restriction does not map directly to quantifying fidelity versus robustness in the quantum domain~\cite{feedbackcontrollaws, statisticalrobustness2}, although it is recovered in some cases~\cite{logsens}. One reason for the discrepancy is that the frequency-domain limitations of classical feedback control have limited applicability in quantifying the performance in the time-domain. The \RIM\ combines the two figures of merit into one single measure: small \RIM\ means high fidelity and high robustness, while large \RIM\ means poor fidelity and poor robustness.

\section{Measuring Robustness and Fidelity of Quantum Controls}\label{sec:method}

\subsection{The General Quantum Control Problem}\label{sec:qcproblem}

The physical system we wish to control is represented by a Hamiltonian
\begin{equation}\label{eq:canoncialcontroldecomp}
  H(t, \mathbf{u}) = H_0 + H_{\mathbf{u}}(t),
\end{equation}
where the time-independent drift Hamiltonian $H_0$ describes the natural dynamics of the system and the control Hamiltonian $H_{\mathbf{u}}(t)$ describes the time-dependent control with the tunable, usually piecewise constant, control parameters $\mathbf{u}$. The closed-system dynamics are governed by the Schr\"odinger equation, which can be written in terms of the unitary evolution operator from time $t_0$ to time $t_1$
\begin{equation}\label{eq:unitaryevoleqn}
  U(t_0,t_1,\mathbf{u}) = \mathcal{T} \exp\mathopen{}\mathclose{\left(-\frac{\imath}{\hbar} \int_{t_0}^{t_1} H(t, \mathbf{u}) \,dt\right)}
\end{equation}
where $\mathcal{T}$ denotes time ordering and $\hbar$ is the reduced Planck constant.

In general, the control problem is formulated as optimizing a fidelity $\mathcal{F}$ over a set of admissible controls. A notion of fidelity that reflects most definitions used in practice is given by $\mathcal{F} := |\left\langle G | K \right\rangle|^2$, which measures the similarity between normalized objects $G$ and {$K$}. If we wish to prepare a state $G = \ket{\psi_f}$ from an initial state $\ket{\psi_0}$ at time $t_0$, then $K = U(t_0,t_1, \mathbf{u})\ket{\psi_0}$ and the optimization problem is given by
\begin{equation}\label{eq:controlprob}
  t_{\text{opt}}, \mathbf{u}_{\text{opt}} = \argmax_{(t_1, \mathbf{u})\in\mathbb{X}} \underbrace{|\bra{\psi_f}U(t_0,t_1, \mathbf{u})\ket{\psi_0}|^2}_{=\mathcal{F}(t_1,\mathbf{u})},
\end{equation}
where $\mathbb{X}$ is the domain of allowed controls, here including the final time $t_1$. A variant, up to normalization, of the state fidelity is the Hilbert-Schmidt inner product $\mathcal{F} = \mathop{\rm Tr}\left(W^\dagger V \right)$ between a desired unitary transformation $W$ and a gate achieved by control, $V = U(t_0,t_1,\mathbf{u})$. In general, we assume the fidelity is bounded, and without loss of generality we assume it lies in $[0,1]$, where $\mathcal{F}=1$ if and only if we have $G=e^{\imath \phi}{K}$, up to a global phase $\phi$ (and equivalently for the Hilbert-Schmidt inner product).

\subsection{Robustness-Infidelity Measure}\label{sec:robustnessmeasure}

Uncertain dynamics turn the fidelity $\mathcal{F}$ into a random variable with a probability distribution $\mathbf{P}(\mathcal{F})$. Intuitively, we call a controller robust if this distribution has a low spread. While a low spread alone may indicate robustness, low fidelity means the controller does not realize the target operation well. So we also expect a fidelity close to $1$. That means the perfect distribution under any uncertainties is $\delta_1$ -- the Dirac delta distribution at maximum fidelity $1$. In particular, we consider the delta function $\delta_x$ to be defined by an indicator cumulative distribution function (CDF),
\begin{equation}\label{eq:deltafunc_cdf}
  C(a) = \begin{cases}
    1 &\quad\text{if } a \geq 0,\\
    0 &\quad\text{if } a < 0.
  \end{cases}
\end{equation}
This permits the familiar delta function property for integration w.r.t. a basic (rapidly diminishing) function,
\begin{equation}
  \int_{-\infty}^\infty g(x)\delta_{x-a}\,dx = \int_{-\infty}^\infty g(x)\,dC(x-a) = g(a).
\end{equation}
Our goal is to define a distance between probability distributions that measures closeness between the ideal and the achieved probability distribution in order to combine high fidelity and its robustness into a single measure.

For this we take the Wasserstein or Earth mover's distance $\mathcal{W}$~\cite{villani, wgan} due to the facts that:
(1) it allows us to compare two probability distributions that do not share a common support, and, in particular, compare discrete and continuous distributions;
(2) its easy geometric interpretation helps with its optimization;
and (3) a simplification allows it to be calculated easily, as shown next.

The dual formulation of the $p$-th order Wasserstein distance~\cite{quantilequiv} between two distributions $\mu$, $\nu$ is given by
\begin{equation}
  \mathcal{W}_p(\mu, \nu) = \sups_{h, g} \left[\int h(x)\,d\mu(x) -\int g(y)\,d\nu(y)\right]^{\frac{1}{p}},
\end{equation}
where $h(x)-g(y) \leq \|x-y\|^p$. Even though this form seems abstract, for one-dimensional distributions, we can analytically compute the optimal maps $h$, $g$ with
\begin{theorem}\label{th:quantileeqiv}(Prop.~1 in~\cite{quantilequiv})
  The $p$-th Wasserstein distance $\mathcal{W}_p(\mu, \nu)$ for one-dimensional probability distributions $\mu$ and $\nu$ with finite $p$-moments can be rewritten as
  \[
    \mathcal{W}_p(\mu, \nu) = \left(\int_0^1 |Q_\mu(z) -Q_\nu(z)|^p \,dz\right)^{\frac{1}{p}}
  \]
  where $Q_\mu(z) = \inf\{x \in \mathbb{R} : C_\mu(x) \geq z\}$ denotes the quantile function and $C_\mu$ is the cumulative probability function of $\mu$ and likewise for $Q_\nu$.
\end{theorem}
Remarkably, the optimal transport distance between one-dimensional distributions $\mu$, $\nu$ over all possible transportation plans can be computed in terms of their quantile functions $Q_{\mu}$, $Q_{\nu}$. From here, following Thm.~\ref{th:quantileeqiv}, it is straightforward to define the $p$-th \emph{Robustness-Infidelity Measure},
\begin{equation}\label{eq:wdfromidealRIMlabel}
  \RIM_p := \mathcal{W}_p(\mathbf{P}(\mathcal{F}), \delta_1) = \left(\int_0^1{|Q_{\mathbf{P}(\mathcal{F})}(z)-1|^p}\,dz\right)^{\frac{1}{p}}.
\end{equation}
It can be written in terms of the raw moments (see App.~\ref{sec:rimcalculations}): {
\begin{equation}
  \RIM_p = \mathds{E}_{f\sim \mathbf{P(\mathcal{F})}}\left[(1-f)^p\right]^{\frac{1}{p}}
\end{equation}
where $f$ is a fidelity sample drawn from the distribution $\mathbf{P}(\mathcal{F})$ and $1-f$ is the corresponding infidelity sample. We use the expectation operator defined as $\mathds{E}_{f\sim \mathbf{P(\mathcal{F})}}[(\cdot)] := \int (\cdot)\mathbf{P}(\mathcal{F}=f)\,df$. For $p=1$, we recover the average infidelity,
\begin{equation}
  \RIM_1 = \mathds{E}_{f\sim \mathbf{P(\mathcal{F})}}\left[1-f\right] = 1 - \mathds{E}_{f\sim \mathbf{P(\mathcal{F})}}\left[f\right].
\end{equation}}

To compute the $\RIM_p$, we estimate $\mathbf{P}(\mathcal{F})$ using $n$ fidelity samples $f_1, f_2, \dotsc, f_n$. Such samples may be obtained in practice via Monte Carlo simulation or physical experiments~\cite{flammia_fid_est}. Hence, barring the computational or experimental expense of obtaining these samples, the $\RIM_p$ is easy to compute. In case the dynamics of the system are certain, i.e.\ $\mathbf{P}(\mathcal{F}) = \delta_f$ for some constant fidelity value $f$, the $\RIM_1$ is equal to the infidelity $1-f$. Moreover, the $\RIM_1$ is small if and only if the controller is robust (in the sense of the fidelity distribution having a low spread) and is also close to the maximum fidelity.

\subsection{The Average Fidelity is Sufficient for Robustness Comparisons}

{We motivate why the $\RIM_1$ is sufficient for comparing robustness and fidelity of controllers by making use of the fact that the \RIM{}s of different orders computed on the estimated fidelity distribution are in agreement. We obtain the following bounds between the lower and higher order \RIM{}s (see App.~\ref{sec:equivalence}):
\begin{subequations}\begin{align}
  \RIM_{p'} &\leq {n}^{\left(\frac{1}{p}-\frac{1}{p'}\right)} \RIM_p, \label{eq:highlow} \\
  \RIM_p    &\leq \RIM_{p'} \label{eq:lowhigh}
\end{align}\end{subequations}
for $p<p'$, where $n$ is the number of samples used to estimate the \RIM. Eq.~\eqref{eq:lowhigh} is stronger and states that $\RIM_p$ is less sensitive to outliers than $\RIM_{p'}$ while Eq.~\eqref{eq:highlow} states that for fixed $n$ and $p$, $\RIM_{p'}$ growth is sublinear ($\propto \exp(-1/p')$). This can be made tighter by adding additional assumptions on the nature of $\mathbf{P}(\mathcal{F})$, but these depend on the specific control problem. The upper bound becomes loose with increasing $n$, but highlights the constraining nature of deviation of higher-order \RIM{}s from $\RIM_1$. 

This means that the higher-order \RIM{}s do not capture more useful robustness information for comparisons, with the base case in Eq.~\eqref{eq:lowhigh} being decided by the $\RIM_1$. $\RIM_1$ has low sensitivity to outliers (see App.~\ref{sec:equivalence}), which makes it easier to estimate than higher-order \RIM{}s, which, like the worst-case fidelity, are harder to accurately practically obtain (as more samples are required, which also explains the presence of $n$ in the inequality).

The bound in Eq.~\eqref{eq:lowhigh} gets tighter for large $n$ but also for decreasing infidelities, so in this regime, the \RIM{}s are in agreement. Another way to see this is to note that the Wasserstein distance provides a structure-preserving geodesic between any fidelity distribution to the ideal $\delta_1$: the distributions converge together with their \RIM{}s of any order. So especially when approaching the ideal distribution $\delta_1$, i.e. in case $\RIM_1$ is small for high-fidelity, robust controller, there is strong agreement between \RIM{}s of all orders. For example, the variance of distributions decreases as $\sim (1-\min \mathcal{F})^2$ as $\min \mathcal{F} \to 1$ in $[0,1]$.

However, the fact that outliers are more influential for higher \RIM\ orders proves useful for optimization~\cite{risksensitivecontrol} where such behavior is sought after, while our goal here is robustness/fidelity comparison. For this goal, in general, outliers are obstructive as they would hide the general distributional trend. From now on we will refer to the $\RIM_1$ without the subscript.

}

\subsection{Perturbations}\label{sec:perturbations}

Next, we define the noise in the system as perturbations of its uncertain dynamics that give rise to $\mathbf{P}(\mathcal{F})$. A perturbation to the full Hamiltonian in Eq.~\eqref{eq:canoncialcontroldecomp} can be expressed as $\tilde{H}(t,\mathbf{u}) = H(t,\mathbf{u}) + \gamma S \in \mathbb{C}^{n \times n}$ where $\gamma \in \mathbb{R}$ describes the strength of a perturbation and $S \in \mathbb{C}^{n \times n}$ its structure, usually normalized using some matrix norm. To induce an uncertainty into the dynamics we treat $\gamma$ and $S$ as random variables drawn from some probability distributions. This give us a general way to represent any physically relevant uncertainties in Hamiltonian parameters.

The structure $S$ may be fixed, e.g., describing the uncertainty in some coupling parameter for the Hamiltonian, while $\gamma$ is drawn from a normal distribution. This would be consistent with a (linear) \emph{structured perturbation} in classical robust control theory~\cite{ssv_mu}. Instead, $S$ may also be drawn from a probability distribution, describing uncertainties across multiple Hamiltonian parameters. While this generalizes structured perturbations, note that they do remain linear w.r.t. the strength. If $S$ is sampled uniformly on the unit-sphere, according to its normalization, we have an \emph{unstructured perturbation}, with (uncertain) strength determined by $\gamma$. Conceptually, if $\gamma$ is drawn from a normal distribution with zero mean and standard deviation $\sigma$, $\gamma S$ describes a ``fuzzy'' ball $\mathcal{B}_\sigma$ around $H(t,\mathbf{u})$. In this paper, we consider unstructured perturbations that are less idealized, in some sense, than the structured perturbations (usually considered in classical control~\cite{doyle2}), allowing the robustness results to be interpreted generically without the need to consider specific sources of uncertainties arising from specific quantum device designs. For simplicity, we write $\mathbf{P}_\sigma(\mathcal{F})$ for a fidelity distribution obtained by unstructured perturbations drawn from $\mathcal{B}_\sigma$.

Our quantification of robustness is dependent on the choice of $\gamma S$ and the uncertainties in these quantities. Note that neither the choice of the noise model nor the magnitude of the noise level is restricted, as our approach is not perturbative around the optimum $(t_{\text{opt}}, {\mathbf{u}}_{\text{opt}})$, which is how noise is usually modelled in the literature~\cite{dayoi, noisefree1, biercuk1, rabitz2, barnes2015robust, rabitzk}. This approach becomes relevant when confidence in an analytical physical model is low or there are missing terms that cannot be analytically or perturbatively accounted for, e.g., complicated noise sources. This is also in accordance with modern robustness theory and the $\mu$ function in classical~\cite{jcdoyle} and quantum~\cite{sophie2021,ssv_mu,cdc21} settings.

\begin{figure}[t]
  \includegraphics[width=\columnwidth]{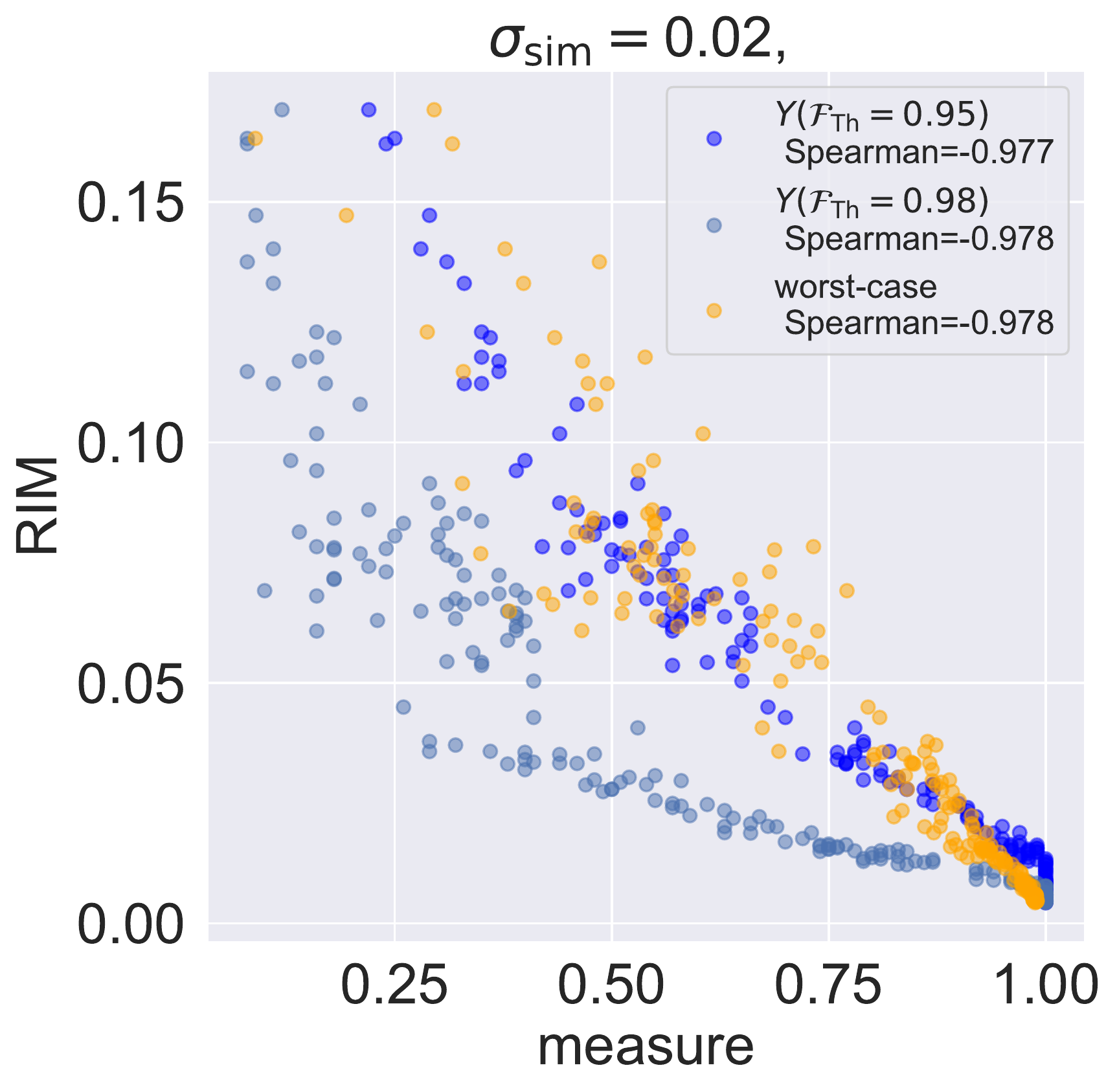}
  \caption{{\RIM\ values generated from $\mathbf{P}_{0.02}(\mathcal{F})$ with $N=100$ samples for $200$ controllers are plotted against the yield $Y(\mathcal{F_{\text{Th}}})$ at fidelity thresholds $\mathcal{F_{\text{Th}}}=0.95, 0.98$ and the worst case fidelity.} Both measures are correlated, as encapsulated by the high negative Spearman correlation coefficients~\cite{spearmanr} and $p$-values $<10^{-4}$.}\label{fig:qfactorintuition}
\end{figure}

To further motivate the \RIM, we study how it compares with other statistical measures of robustness. The \RIM\ generally correlates with { worst-case or minimum sample fidelity, variance or higher moments and the yield function $Y(\mathcal{F_{\text{Th}}})$, which is the fraction of fidelities greater than a threshold fidelity $\mathcal{F_{\text{Th}}}$. Fig.~\ref{fig:qfactorintuition} shows a scatter plot of \RIM\ values versus $Y(0.95)$, $Y(0.98)$ and the worst-case fidelity} for an example problem using Eq.~\eqref{eq:firstexcitationsubspaceXX} discussed in Sec.~\ref{sec:problem}. The \RIM\ has an advantage over $Y$ in that it does not depend on an arbitrary choice of $\mathcal{F}_{\text{Th}}$.

\subsection{Measuring the Performance of Control Algorithms}

We can also apply the previous arguments to derive a measure to compare the ability of control algorithms to find high-fidelity, robust controllers. Let $\mathbf{P}(\RIM)$ be a distribution of \RIM\ values of controllers obtained by a particular algorithm and a particular control problem with specific uncertainties. This can be estimated by sampling $L$ controllers produced by the algorithm. The ideal of this distribution is $\delta_0$, so that we can define the \emph{Algorithmic Robustness Infidelity Measure},
\begin{equation}\label{eq:arim}
  \ARIM := \mathcal{W}_1(\mathbf{P}(\RIM), \delta_0)
        = \mathds{E}_{r\sim \mathbf{P}(\RIM)}\left[r\right],
\end{equation}
following the same argument as before. The \ARIM\ is small if and only if the underlying \RIM\ distribution $\mathbf{P}(\RIM)$ has higher density at or near $\RIM=0$, i.e. is close to the ideal $\delta_0$.

\section{Robustness for Static Control Problems}

We study the robustness of static control problems, where the controls are time independent, instead of the usual time dependent controls. Previous work has shown that particularly robust controls can be found for these systems~\cite{feedbackcontrollaws, statisticalrobustness1, statisticalrobustness2, ssv_mu}. While these systems are often not fully controllable, solutions for specific operations can be found via optimization~\cite{time_optimal_cdc}. The static approach is simpler in the sense of having fewer control parameters to optimize over, which reduces computational and experimental complexity. This makes the problem suitable to demonstrate the practical usage of the \RIM\ in a concrete example and explore the robustness properties of the control algorithms as well as the controllers they find.

\subsection{Information Transfer in the Single Excitation Subspace of XX Spin Chains}\label{sec:problem}

We consider a network of $M$ spins represented by the quantum Heisenberg model given by the Hamiltonian
\begin{equation}\label{eq:Heisenberg}
  \frac{H_\text{heis}}{\hbar} = \sum_{a \in \{x,y,z\}}{{\sum_{j=1}^M J^a \sigma_j^a\sigma_{j+1}^a}} + \eta \sum_{j=1}^M{\sigma^z_j}
\end{equation}
where $\sigma_j^a=\mathds{I}^{\otimes j-1}\otimes \sigma^a \otimes \mathds{I}^{\otimes M-j}$ and $\sigma^a$ are the usual Pauli matrices. We set $J^z= 0$ and $J^x=J^y=J$ for the XX model with uniform couplings. This model has been studied extensively, starting with Ref.~\cite{Lieb} in 1961, and a more recent review of the system, as it relates to quantum communication, is provided in Ref.~\cite{Bose}. Conditions for perfect state transfer along XX chains were derived in Ref.~\cite{Christandl} and applied to NMR systems~\cite{Lu}. Similar experiments have been carried out in photonic systems~\cite{Tzortzakis, Christodoulides}, and proposals for engineering similar systems with trapped ions~\cite{Lewenstein} and cold atoms~\cite{Fallani} exist.

The state space naturally decomposes into non-inter\-acting excitation subspaces as the Hamiltonian commutes with the total excitation operator. Here we consider the first excitation subspace, the smallest space that enables transfer of one bit of information between the nodes in the network. Higher excitation subspaces may be needed for other applications, but it is desirable for information transfer to limit the space to the smallest space that is sufficient to achieve the task. This is a much smaller space and only grows as $O(M^2)$ as opposed to $O(\exp(2M))$. The Hamiltonian of the first excitation subspace is
\begin{equation}\label{eq:firstexcitationsubspaceXX}
  \frac{(H_{XX})_{l,m}}{\hbar} = J \mathds{1}_{l,m\pm 1} + \Delta_l\mathds{1}_{l,m}
\end{equation}
where $\mathds{1}_{l,m}$ is the Kronecker delta. The static controls are local energy biases $\Delta_l$ on spin $\ket{l}$ in a diagonal matrix $H_\mathbf{\Delta} = \mathop{diag}(\Delta_1,\dotsc,\Delta_M)$.

\begin{figure}[t]
  \centerline{\includegraphics[width=\columnwidth]{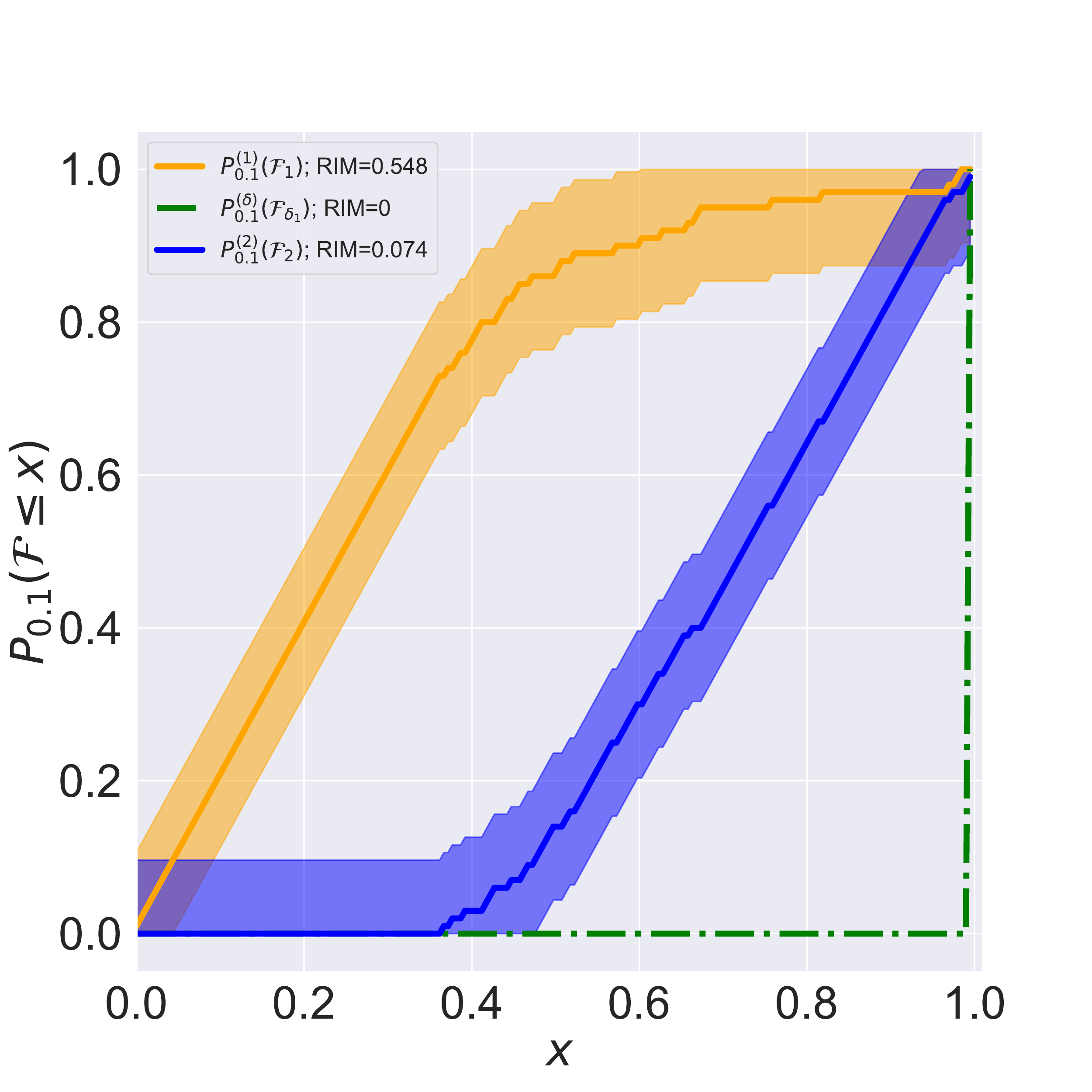}}
  \caption{To illustrate the \RIM\ robustness measure, two static controllers for an XX spin chain of length five for transferring an excitation from spin $\ket{1}$ to $\ket{3}$ are compared. The empirical approximations to the CDFs for the two controllers, $l=1,2$, were simulated using $100$ bootstrapped perturbations with $\sigma=0.1$, giving fidelity distributions $\mathbf{P}_{0.1}(\mathcal{F}_l)$ for the fidelity random variables $\mathcal{F}_l$. The fidelity distribution $\mathbf{P}_{0.1}(\mathcal{F}_{\delta_1})$ for a perfectly robust controller with $\mathcal{F}_{\delta_1}$ is also shown. The ECDFs are estimated using $500$ bootstrap repetitions. The $0.95$ confidence bounds on their error are obtained using the Dvoretsky-Kiefer-Wolfowitz inequality~\cite{dkwplug}. Closeness to the perfectly robust controller can be interpreted as having a smaller area under the curve and is indicated by the \RIM\ values.}\label{fig:qexample}
\end{figure}

$H_{XX}$ allows for transfer of single bit excitations from an initial spin state $\ket{a}$ to a final state $\ket{b}$. We define the fidelity as $\mathcal{F}=|\bra{b}U(t_0,t_1,H_{XX})\ket{a}|^2$ and the infidelity as $\mathcal{I}=1-\mathcal{F}$. The solution to Eq.~\eqref{eq:controlprob} is a final time $t_{\text{opt}}$ and a single vector of $M$ biases $\mathbf{\Delta}_{\text{opt}}$. The perturbations are given by
\begin{equation}\label{eq:perturbation}
  \left(S_\sigma\right)_{l,m} = \sum_{k=1}^{M-1} \gamma_k^J J \mathds{1}_{l,k} \mathds{1}_{l,m\pm 1} + \sum_{c=1}^M \gamma_c^C \Delta_c \mathds{1}_{c,l} \mathds{1}_{l,m}
\end{equation}
where $\gamma_k^J$ and $\gamma_c^C$ are the strength of the perturbation on the couplings and controls respectively. We draw these strengths from the same normal distribution $\mathcal{N}\left(0,\sigma^2\right)$ with mean $0$ and variance $\sigma^2$. A numerical example illustrating the \RIM\ via the empirical CDF (ECDF) for two controllers is shown in Fig.~\ref{fig:qexample}.

Depending on the hardware platform, it is possible to consider specific practically motivated correlated noise models with correlated structured perturbations or a power law decaying electric-field noise ($1/s$), e.g., in trapped atomic platforms~\cite{rev3,1/snoise}. We have chosen to implement the simplest option of equal strength random perturbations on all non-zero entries of the Hamiltonian that is also relevant in practical settings~\cite{Fallani,Tzortzakis,Christodoulides,Lewenstein,Lu}.

\subsection{Algorithms for Static Control Problems}\label{sec:algomotivation}

The \ARIM\ compares algorithm performance in finding robust controllers. Here, we describe the algorithms used to find the controllers for the static control problems. For selecting algorithms, we tried to (a) investigate the performance of algorithms commonly used in the quantum control {and other communities}, (b) consider algorithms that do and do not require gradient information, and (c) consider reinforcement learning, more recently also used in quantum control.

L-BFGS is a common optimization algorithm used in quantum control as part of GRAPE~\cite{grape} and performed well on finding high-fidelity energy landscape controllers~\cite{time_optimal_cdc}. It has not been designed for noisy optimization, but there exist smoothing modifications that attempt to address this~\cite{lbfgssmoothing1, sophie_lbfgssmoothing, noisylbfgs}. For individual controller comparisons, we use standard L-BFGS with an ordinary differential equation model to compute the fidelity without perturbations during optimization. This serves as a baseline to understand the performance of optimizing noiseless objective functionals compared to the noisy optimization performed by all other selected algorithms and its impact on the robustness of the controllers found. We have explored stochastic gradient descent methods (e.g.\ ADAM~\cite{adam}) and also tested a noisy version of L-BFGS that has been recently proposed that modifies the line search and lengthening procedure during the gradient update step~\cite{noisylbfgs} and found that our training noise scales were too large and washed away gradient information, rendering these algorithms unsuitable for our study.

Reinforcement learning has been successfully used for tackling quantum control in challenging noisy environments, resulting in similar or better performances compared to standard control methods. Promising results include the stabilization of a particle via feedback in an unstable potential~\cite{qrl_cartpole}, optimizing circuit-QED, two-qubit unitary operators under physical realization constraints~\cite{qrl_alphazero}, and optimizing multi-qubit control landscapes suffering from control leakage and stochastic model errors~\cite{qrl_niu}, among many others.

Proximal Policy Optimization (PPO) is a policy gradient method in the class of reinforcement learning algorithms~\cite{barto}. It uses a discounted reward signal (e.g., the fidelity) accumulated over multiple interactions with the optimization landscape using non-parametric models: the policy function, that does the interacting by performing control operations, and the action value function, that predicts the quality of each action undertaken by the policy in terms of future payoffs in the reward signal. Both are estimated using neural networks in a control problem agnostic fashion. Doing this allows the incorporation of perturbations during training which specifically has advantages in finding robust controls for energy landscape problems~\cite{self1}. Here, we use a control problem formulation of PPO for Eq.~\eqref{eq:firstexcitationsubspaceXX} as described in Ref.~\cite{self1}.

Nelder-Mead is a popular simplex-based control algorithm using direct search. Essentially, it keeps updating a polytope whose vertices are function evaluations towards an optimum direction. It has successfully been used in noisy experimental settings~\cite{Weidner_PRL} due to its non-reliance on gradient information~\cite{nmsell1, nmsell2, nmsell3}, especially when obtaining such information is resource-intensive.

Stable Noisy Optimization by Branch and Fit (SNOBFit) has been chosen as it has been designed to filter out quite large scale noise in objective functionals~\cite{snob}. It fits local models using objective function evaluations and implements a branching and splitting algorithm to partition the parameter space into smaller boxes with one function evaluation per box. The latter is a non-local search scheme that orders promising sub-boxes by the number of bisections required to get from the base box to that box. Sub-boxes with smaller bisections are worth exploring more. Like PPO, it does not rely on explicit gradient information and builds models of the optimization landscape. Thus, both algorithms should be able to cope with large amounts of noise in the form of controller and model uncertainties, environmental effects and singularity during optimization. SNOBFit, however, differs importantly from PPO in the assumption that those models are linear. Moreover, its non-local optimization landscape exploration is not random and thus has comparatively a lot less variance in performance (that may or may not be poor).

\section{Numerical Experiments}\label{sec:numexp}

\begin{figure*}[t]
  \includegraphics[width=0.64\textwidth]{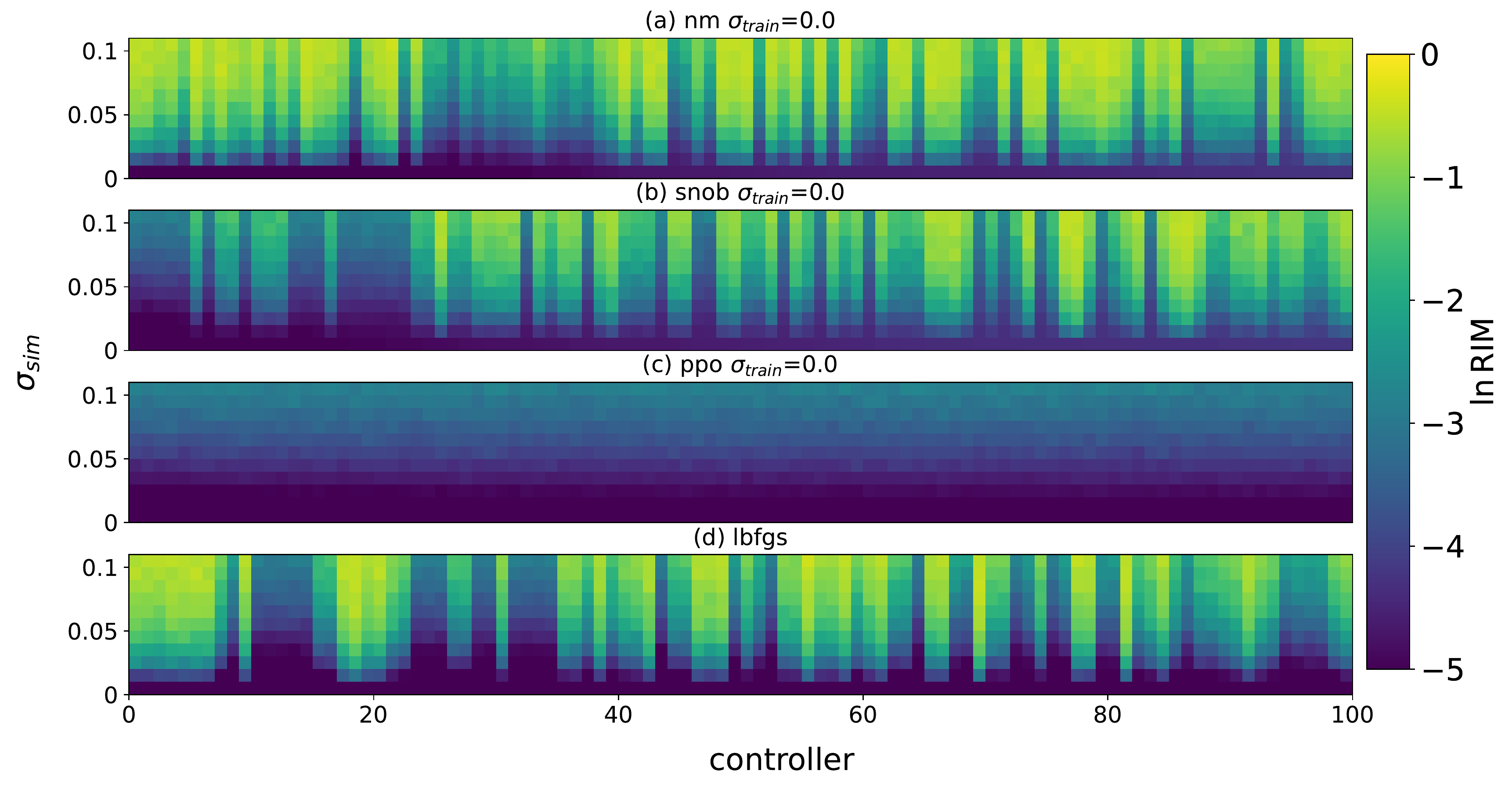}
  \hfill
  \includegraphics[width=.35\textwidth]{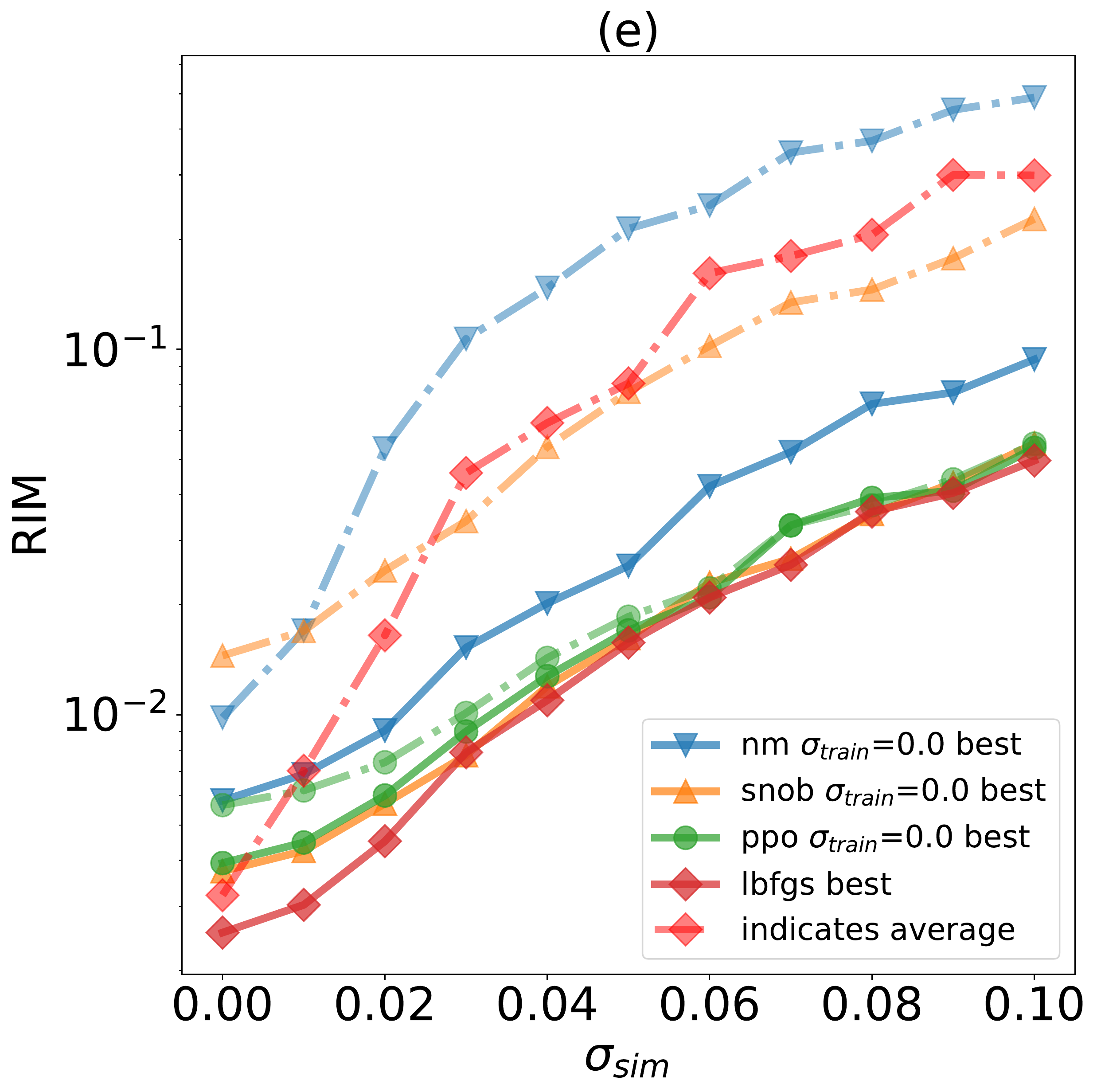}\\
  \caption{(a)-(d) $100$ controllers found for the XX spin chain model, Eq.~\eqref{eq:firstexcitationsubspaceXX}, using Nelder-Mead, SNOBFit, PPO, and L-BFGS for $M=5$ and the spin transition from $\ket{1}$ to $\ket{3}$ with $\sigma_\text{train}=0$. The controllers are ranked in increasing order of infidelity at $\sigma_\text{sim}=0$ from left to right. Each column represents a single controller's \RIM\ at $\sigma_\text{sim} = 0, 0.01, 0.02, \dotsc, 0.1$ from the bottom to the top on a log scale. Even if the infidelity or \RIM\ at $\sigma_\text{sim}=0$ is close to $0$, some controllers' \RIM\ values degrade faster than others and are hence less robust despite starting at very low infidelities.
  (e) \RIM\ as a function of $\sigma_\text{sim}$ for the average and best controller (i.e., most dark over all $\sigma_\text{sim}$ levels) out of the $100$ shown in (a)-(d) in terms of how much they preserve their corresponding \RIM\ rank average across all $\sigma_\text{sim}$. Each algorithm is indicated by a marker shape, and the solid and dash-dotted lines denote the best and average controller lines respectively. All the best controllers have very high initial fidelities and are very similar across the different control algorithms, with Nelder-Mead being only moderately worse.}\label{fig:noiselessindividualconts}
\end{figure*}

To explore the robustness of controllers and corresponding control algorithms (see experimental motivation in Sec.~\ref{sec:intro}), we perform a Monte Carlo robustness analysis using the \RIM\ on numerical solutions to the spin chain information transfer problem in Sec.~\ref{sec:problem} for chains of length $M=4,5,6,7,8,9$ with $J=1$.

We look at transitions from the start of the chain $\ket{1}$ to the end $\ket{M}$ and from $\ket{1}$ to the middle $\ket{\ceil{\tfrac{M}{2}}}$. The former transition is physically easy to control while the latter is more challenging~\cite{time_optimal_cdc} as transitions to the middle exhibit anti-core behavior~\cite{anticore}.

We collect the best $100$ solutions, ranked by their fidelity, obtained by all the control algorithms. Each algorithm has a budget of $10^6$ fidelity function evaluations. The budget correlates with the run time for each algorithm. It is imposed to allow for a fair comparison of the algorithm robustness performance under similar resources, while being agnostic to specific implementations and speed differences.

We initialize $\mathbf{\Delta}, t$ with quasi Monte Carlo samples from the Latin Hypercube~\cite{latin1, latin2, quasimontecarlo} to increase convergence rate and decrease clustering of controllers. This permits coverage of the parameter domain with $O(1/\sqrt{N})$ samples as opposed to $O(1/N)$ for random sampling, where $N$ is the number of initial values. Our constraints are $0\le t_f\le 70$ and $-10\le \Delta \le 10$. We use $100$ bootstrap samples to estimate fidelity distributions throughout. The perturbation strengths $\gamma_j^J$ and $\gamma_c^C$ are scaled by $J$ and $\Delta$ respectively as per Eq.~\eqref{eq:perturbation}. Note, for $\sigma=0$, $\mathbf{P}_{0}(\mathcal{F}) = \delta_{\cal F}$ is deterministic.

The perturbation strengths are drawn from a normal distribution with standard deviation $\sigma_\text{train}$ determining the strength of the noise added for the optimization. $\sigma_\text{sim}$ is the noise level used in the simulations to assess the robustness of the controllers found. { Implementation details are in App.~\ref{sec:optobjectives}.} The optimization objective is noiseless $\mathcal{F}$ for Sec.~\ref{sec:individualcont}, stochastic $\mathcal{F}$ with unstructured perturbation $S_{\sigma}$ for Sec.~\ref{sec:algocomparison}, and the \RIM\ for the non-stochastic problem and a stochastic $\mathcal{F}$ with $S_{\sigma}$ in Sec.~\ref{sec:uncons_algocomparison}. 

\subsection{Characterization of All Controllers Found with Constrained Resources}\label{sec:individualcont}

\subsubsection{Ranking Individual Controllers}\label{sec:nonoise_contcomp}

In this section, we address our motivating question (A), whether high fidelity implies high robustness for an individual controller. We also numerically demonstrate the non-linear and non-uniform deterioration of robustness with increasing noise which implies a trade-off between higher fidelity at no noise and robustness at higher noise levels.

To this end, we employ control algorithms to optimize an objective functional without noise, i.e., setting $\sigma_\text{train}=0$ (see Sec.~\ref{sec:perturbations}), under the general optimization conditions outlined at the start of Sec.~\ref{sec:numexp}. We rank these controllers by their infidelity values and then compute the \RIM\ values for various levels of simulation noise, $\sigma_\text{sim} = 0.01, 0.02, \dotsc, 0.1$.

For example, Figs.~\ref{fig:noiselessindividualconts}(a)-(d) show a pseudo-color plot of the \RIM\ values for $100$ controllers found for the chosen test control problem (chain of length $M=5$, target spin transfer $\ket{1}$ to $\ket{3}$). The lowest infidelity controllers start from the left and are indexed by columns $1$ to $100$ indicating their respective ranks according to their \RIM\ at $\sigma_\text{sim} = 0$. The \RIM\ values, as a function of $\sigma_\text{sim}$, for individual controllers grow at different rates despite starting at quite similar small values for all algorithms. The main result, that applies also to all transitions (not explicitly shown here), is that the high fidelity controllers do not, in general, preserve their ranks as $\sigma_\text{sim}$ increases. E.g., for SNOBFit (see Fig.~\ref{fig:noiselessindividualconts}(b)), the \RIM\ for controllers $6, 8, 9, 11--13$ grows much more rapidly than for controllers $24--33$, indicated by rapid color changes from dark (low \RIM) to light (high \RIM) in the vertical direction. Interestingly, almost all controllers found by PPO have very low \RIM\ across $\sigma_\text{sim}$ values compared to the other control algorithms (color remains dark for longer). This is, however, not reflective of PPO's general behavior on the extended sample of problems we examined (see App.~\ref{sec:arim_allspins}). It could be limited fundamentally by the existence of robust controllers and/or the resource budget for a particular problem (see Fig.~\ref{fig:noiselessindividualcontssupp} in App.~\ref{sec:extraplotscontcomp} showing results for other transitions).

We further evaluate the best performing individual controller. To this end, we seek the controller that preserves its overall \RIM\ rank average the most across the noise levels. It is computed using the reshuffled \RIM\ ranks of each controller for all values of $\sigma_\text{sim}$. Likewise, we locate the controller that has the median \RIM\ rank average across the noise levels as the averagely performing controller. Most of the \RIM\ rank sum distributions studied were symmetric, and their median was close to their average value. So we can try to understand average controller \RIM\ rank order consistency in terms of how the median controller performs. We compare the \RIM\ values of the median with the best controller in Fig.~\ref{fig:noiselessindividualconts}(e) for all algorithms, showing the $\RIM$ values for the best and median controller as a function of $\sigma_\text{sim}$.

For all algorithms, the best and the average controllers have similar infidelities (initial \RIM\ value) in Fig.~\ref{fig:noiselessindividualconts}(e). Their behavior as a function of $\sigma_{\text{sim}}$ is different and is generally non-linear. Thus, the best controllers, despite being distinguishable from the others at $\sigma_{\text{sim}} = 0$, become indistinguishable for higher $\sigma_{\text{sim}}$ and point at a trade-off between infidelity (at no noise) and robustness that could be leveraged when selecting a controller to be deployed for a noisy system. Moreover, the \RIM\ curve of the best controller among all algorithms (here L-BFGS) suggests a fundamental limitation on \RIM\ for this problem. It is likely not possible to obtain curves that are lower, but this remains theoretically unresolved.

\subsubsection{Ordinal Kendall Tau for ${\rm RIM}_{\sigma_\text{sim}}$-Rank Consistency}\label{sec:consistencystat}

To address the motivating question (B), how likely a given algorithm is to produce robust controllers that were obtained in an ideal (no-noise) setting, we are interested in how consistently a controller acquisition strategy produces controllers with low \RIM.

To that end, we reduce the \RIM\ rank consistency property of the top-$k$ controllers across two perturbation strengths $\sigma_\text{sim}^{(i)}$ and $\sigma_\text{sim}^{(j)}$ to a prediction problem by asking the following: \textbf{(Q)} \emph{How well does the \RIM\ rank of a controller, when ordered at strength $\sigma_\text{sim}^{(i)}$, predict the \RIM\ rank of the controller at strength $\sigma_\text{sim}^{(j)}$?} 

To answer this question, let us denote the controller \RIM\ $\sigma_{\text{sim}}^{(i)}$-rank order by the vector $\mathbf{r}^{\sigma^{(i)}_\text{sim}}$, and compute an ordinal (binned/categorical) version of the Kendall-tau-B statistic $\tilde{\tau}$~\cite{kendalltau1, kendalltaub}, a measure of statistical dependence between $\mathbf{r}^{\sigma^{(i)}_\text{sim}}$ and $\mathbf{r}^{\sigma^{(j)}_\text{sim}}$. The ordinals are constructed only for $\mathbf{r}^{\sigma^{(i)}_\text{sim}}$ by binning using a discrepancy parameter $\alpha$ that indicates the fraction of the maximum \RIM\ value difference within a single bin. The binned rank order $\tilde{\mathbf{r}}^{\sigma^{(i)}_\text{sim}}(\alpha)$ minimizes the effect of small movement in either rank due to noise. Then $\tilde{\tau}$ is computed by
\begin{equation}\label{eq:ordinaltau}
  \tilde{\tau}(\sigma_{\text{sim}}^{(i)}, \sigma_{\text{sim}}^{(j)}) = \tilde{\tau}_{i,j} =
  \frac{\sum_{l<m}\mathds{I}^+_{l,m}+\mathds{I}^-_{l,m}}{\sqrt{\left(K-t_\text{total}^{(i)}\right) \left(K-t_\text{total}^{(j)}\right)}}
\end{equation}
where
\begin{equation}\mathds{I}_{l,m} =
    \mathop{sgn}\left(\tilde{\mathbf{r}}^{\sigma^{(i)}_\text{sim}}_l - \tilde{\mathbf{r}}^{\sigma^{(i)}_\text{sim}}_m\right)
    \mathop{sgn}\left(\mathbf{r}^{\sigma^{(j)}_\text{sim}}_l - \mathbf{r}^{\sigma^{(j)}_\text{sim}}_m\right)
\end{equation}
are the $l,m$-th sign products of the rank order differences at $\sigma^{(i)}_\text{sim},\sigma^{(j)}_\text{sim}$ with $+/-$ denoting the positive/negative pair contributions; $K=k(k-1)/2$ is the number of total pairs being compared; $t_\text{total}^{(i)} = \sum_l{t_l^{\sigma^{(i)}_\text{sim}}(t_l^{\sigma^{(i)}_\text{sim}}-1)/2}$ are the total numbers of ties where $\mathds{I}_{l,m} = 0$ for $\sigma^{(i)}_\text{sim}$ and likewise for $t_\text{total}^{(j)}$. For complete positive/negative rank order correlation $\tilde{\tau}=\pm 1$ and $\tilde{\tau} = 0$ for zero rank order correlation. For our hypothesis test, we assumed a worst case $p$-value of $10^{-4}$ as an acceptance criterion on the numerical results that will follow and also that the controllers generating these rank orders are independent of each other. In this case, this constraint is satisfied by the i.i.d. noise model for a given set of unique controllers corresponding to different points in a static optimization landscape. The independence over the choice of controllers is not necessary as all the consistency comparisons are for this fixed choice of controllers.

\begin{figure}[t]
  \includegraphics[width=\columnwidth]{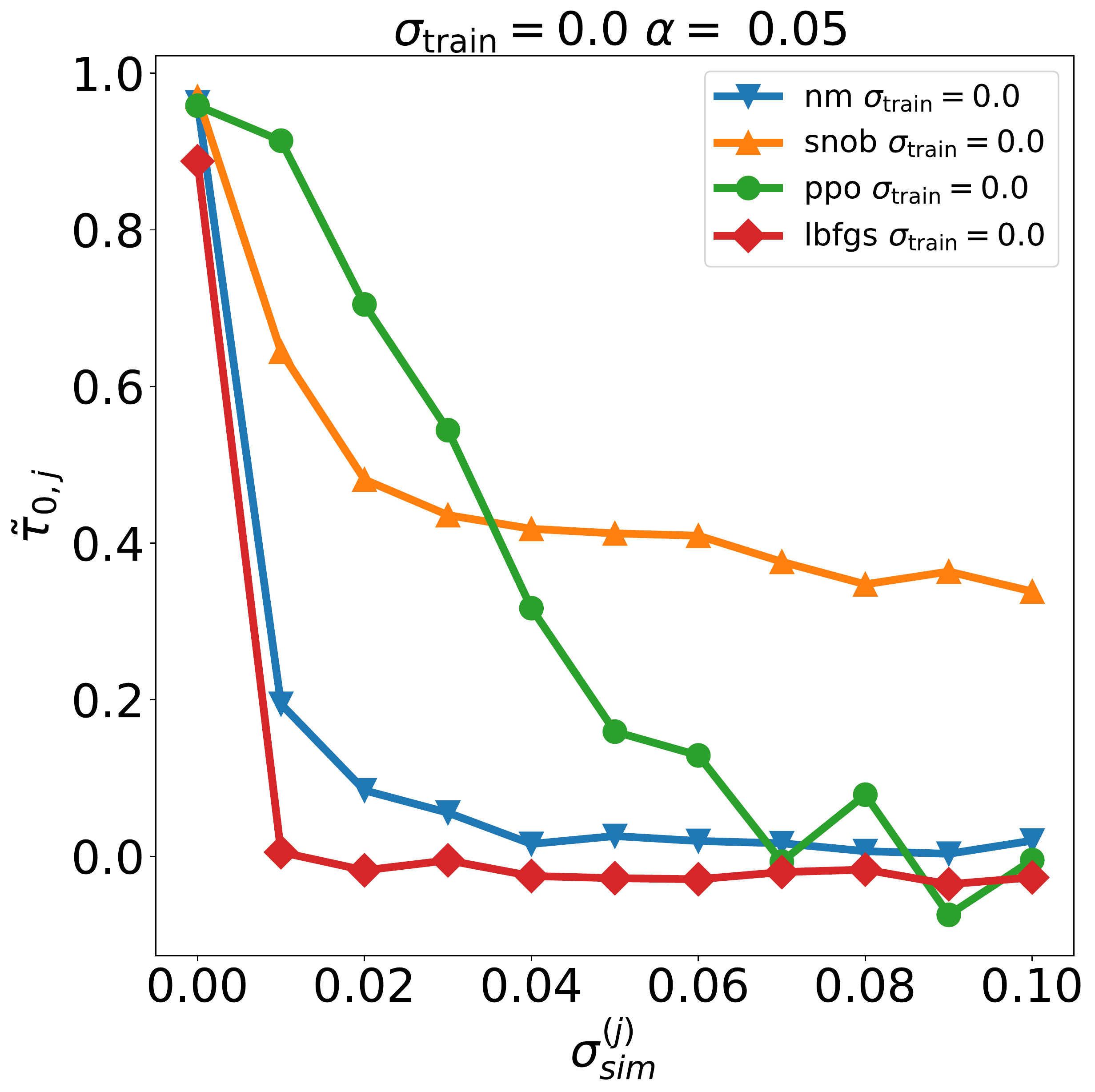}
  \caption{\RIM\ rank order consistency statistic $\tilde{\tau}$ for the $100$ controllers found for the problem $M=5, \ket{1}$ to $\ket{3}$ between the two levels: no simulation noise, $\sigma_\text{sim}^{(i)}=0$ and $\sigma_\text{sim}^{(j)}$ from $\{0.0, 0.01, \dotsc, 0.1\}$ for (a) Nelder-Mead, (b) SNOBFit, (c) PPO, and (d) L-BFGS without training noise. In other words, this is the correlation of infidelity rank order with the general \RIM\ ranks. The $\tilde{\tau}_{0,j}$ values decline the slowest for PPO until $\sigma_\text{sim}^{(j)}=0.04$ and then SNOBFit takes over compared to the rest. This shows, for this case, that the PPO infidelity rank order correlates the most with \RIM\ rank order for $\sigma_\text{sim} \leq 0.03$.}\label{fig:taumatrix}
\end{figure}

\begin{figure*}[t]
  \includegraphics[width=0.47\textwidth]{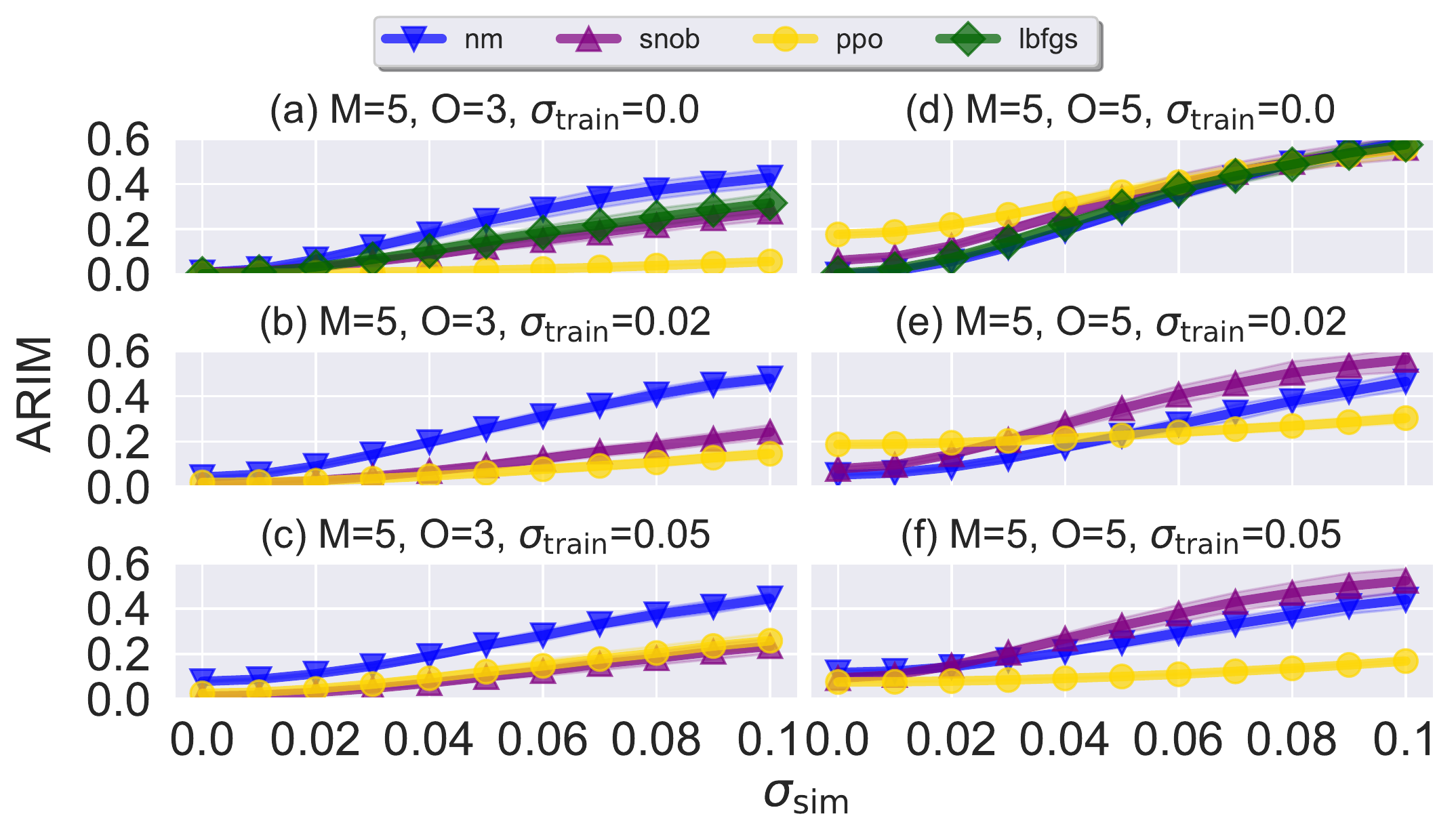}
  \hfill
  \includegraphics[width=.52\textwidth]{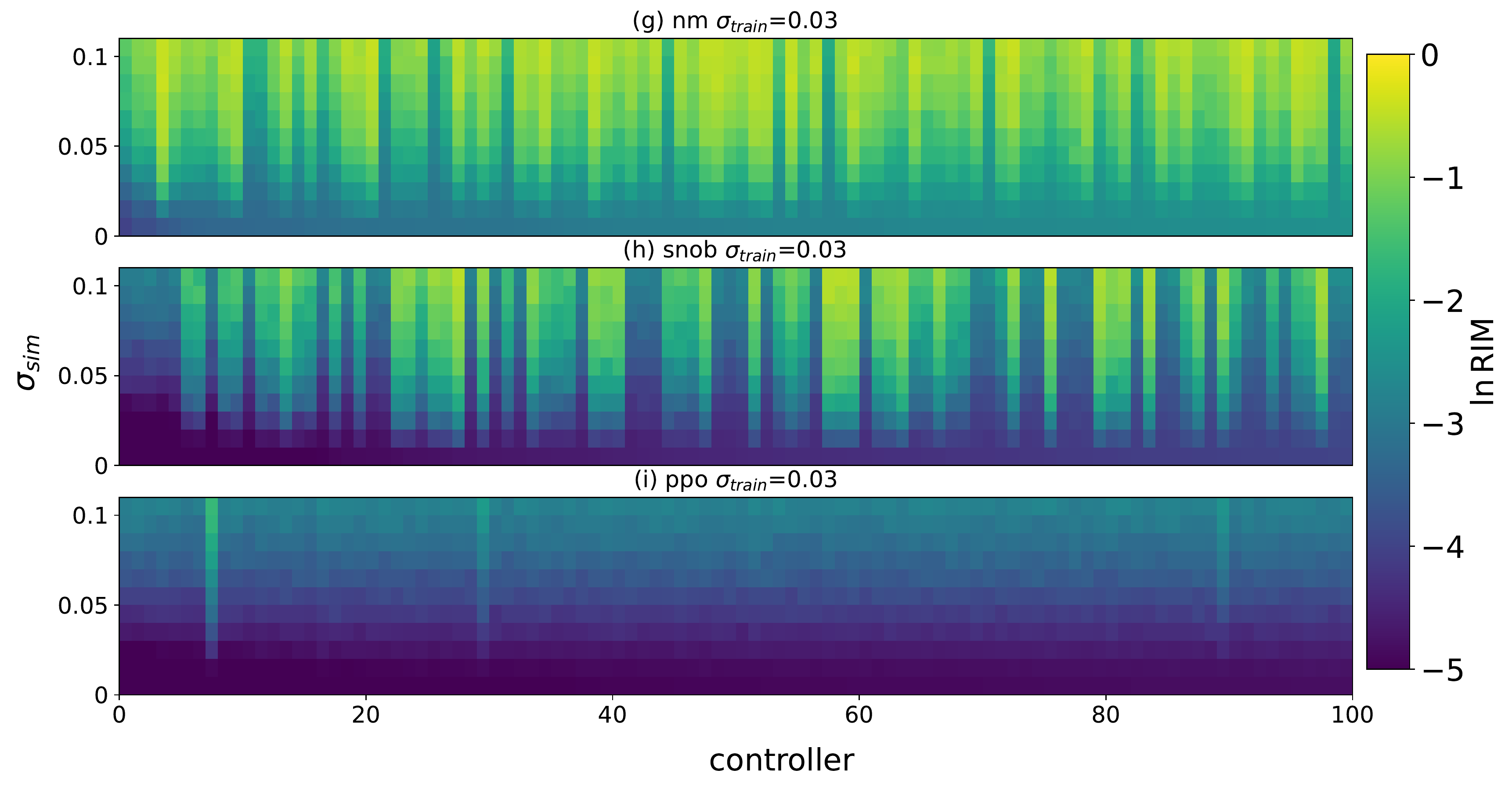}\\
  \caption{
  {
  \ARIM\ as a function of $\sigma_\text{sim}$ for $M=5$ where (a)-(c) are the end-to-middle ($\ket{1}$ to $\ket{3}$) and (d)-(f) are the end-to-end ($\ket{1}$ to $\ket{5}$) transitions (end denoted by $O$). The \ARIM\ is computed from a distribution of \RIM\ values for $100$ controllers for each $\sigma_\text{sim}$ for SNOBFit, Nelder-Mead, PPO and L-BFGS. We identify each algorithm with a unique marker and/or color. Both PPO and SNOBFit are run multiple times at $\sigma_\text{train}=0,0.02,0.05$. PPO has higher variance w.r.t.\ $\sigma_\text{train}$ than SNOBFit and Nelder-Mead, whose performance curves are more in line with the L-BFGS curve for $\sigma_\text{sim} \geq 0.05$ and mostly worse for $\sigma_\text{sim} \leq 0.05$. 95\% confidence intervals (shading) are computed using non-parametric bootstrap resampling~\cite{bootstrapresampling} with $100$ resamples.
  (g)-(i) show individual-controller comparisons for $\sigma_\text{train}=0.03$ ranked by fidelity (leftmost is highest).
  }}\label{fig:arimall}
\end{figure*}

For our earlier spin chain example ($M=5$ spins, transfer from $\ket{1}$ to $\ket{3}$), we focus on $\tilde{\tau}$ for $\sigma^{(i)}_\text{sim}=0, \sigma^{(j)}_\text{sim}$ pairs that is sufficient to answer \textbf{(Q)}. More specifically, we aim to understand how well the no-noise \RIM\ (i.e., the average infidelity) ranks correlate with the general \RIM\ ranks. This is shown in Fig.~\ref{fig:taumatrix} for each optimization algorithm for $\alpha=0.05$. For the $\sigma^{(i)}_\text{sim} \geq 0.03$, the \RIM\ rank order is the most consistent with $\tilde{\tau} \gtrsim 0.6$ for PPO excluding other algorithms. But there is larger shuffling of the ranks of PPO controllers as $\sigma_\text{sim}$ increases with deteriorating $\tilde{\tau}$ and SNOBFit takes over. This may be due to small numerical differences in \RIM\ (see Fig.~\ref{fig:noiselessindividualconts}(c)) observed, and thus a stronger consistency for $\sigma_\text{sim} \leq 0.03$ is captured. We highlight in App.~\ref{sec:ppooptrim} that the reason why PPO infidelities correlate more with \RIM\ values at higher $\sigma_\text{sim}$ is because it optimizes a discounted \RIM\ ($\sum_i{\gamma^i{\RIM^{(i)}}}$ for $0 \leq \gamma \leq 1$) as its reward function. We extend this analysis for $\sigma_\text{train} > 0$ in App.~\ref{sec:moreconst_stat_pppo} to further corroborate that the infidelity rank order for PPO correlates most with higher order \RIM{}s.

The other algorithms typically have a sharper drop at $\sigma^{j)}_\text{sim}=0, 0.01$ step where the infidelity rank order for L-BFGS and, to a lesser extent, Nelder-Mead is completely non-informative (due to very high fidelity values without noise) and is not consistent with the orders at larger $\sigma_\text{sim}$. This is most likely because the controllers found are the result of second order, gradient-based or similarly successful search methods for finding optima precisely. Since PPO and SNOBFit are gradient-free, for $\sigma_\text{sim} \geq 0.03$, their controllers are more consistent in comparison. In this case, the infidelity rank order is more informative of the \RIM\ rank order than, e.g., L-BFGS, as fidelities are not being fully maximized due to the absence of a strong gradient direction. Note that a viable link between the consistency statistic and a generic gradient-based algorithm is hard to establish, so this does not preclude the existence of algorithms that are $\tilde{\tau}$-wise better.

Finally, note that $\tilde{\tau}$ should be thought of as a proxy of reliability of an algorithm's capability to generate numerical control solutions whose infidelity values are more consistent and predictive of their \RIM\ values at higher $\sigma_\text{sim}$. If strong correlation is obtained, this circumvents (or at least increases confidence for circumventing the latter's) computation.

However, high \RIM\ rank order consistency does not imply that the \RIM\ values remain low at higher noise. Rather, it indicates how much the \RIM\ of a controller is predictive of the controller's relative robustness performance at a higher noise level. The non-parametric nature of $\tilde{\tau}$ removes information about the fidelity value range and should be viewed in conjunction with Fig.~$\ref{fig:noiselessindividualconts}(a)-(e)$. If the correlation signal is strong, it could be used to sidestep the evaluation of the \RIM\ at non-zero noise in favor of using the infidelity instead, eliminating the need for expensive sampling.

\subsection{Comparison of Control Algorithms with Constrained Resources}\label{sec:algocomparison}

We address our motivating question (C): what is the effect of training noise on a control algorithm's ability to find robust controllers? The overall picture is complex in terms of algorithm rankings. We numerically confirm that there is a problem-dependent optimal noise level that best smooths the optimization landscape for algorithms to more consistently find robust controllers.

We collect $100$ controllers at training perturbations $S_{\sigma_{\rm train}}$ with training noise level $\sigma_\text{train} \in \{0, 0.01, \dotsc, 0.05\}$ for PPO, SNOBFit and Nelder-Mead. We do not consider any training noise for L-BFGS, since only the former algorithms are designed to perform optimization with noisy perturbations. This involves using a stochastic fidelity (objective) function call evaluated under the single structured perturbation $S_{\sigma_\text{train}}$ (exactly analogous to $S_{\sigma_\text{sim}}$).

We select $\sigma_\text{sim} \in \{0, 0.01, \dotsc, 0.1\}$ to evaluate the \RIM\ of the controllers found at different noise levels with a budget of $10^6$ objective function calls per run. Each run corresponds to $100$ controllers found under this budget constraint. The \ARIM\ is then used to quantify an algorithm's performance, w.r.t. robustness and fidelity, based on the $100$ controllers that it found during the run.

We only show the representative end-to-middle and end-to-end transition for the state-preparation problem for $M=5$ at $\sigma_\text{train} = 0, 0.02, 0.05$ in Fig.~\ref{fig:arimall}. Results for other spin-transitions and training noises are presented in App.~\ref{sec:arim_allspins}.

Recall from Eq.~\eqref{eq:arim} that the \ARIM\ is a measure of how far the distribution $\mathbf{P}(\RIM)$ is from its ideal $\delta_0$. The \ARIM\ curves at different training noises in Fig.~\ref{fig:arimall}(a-f) increase at different rates $\sigma_\text{sim}$, starting from similar base \ARIM\ values at $\sigma_\text{sim}=0$ for each algorithm. Note that the base \ARIM\ value coincides with the average infidelity over controllers, in the absence of training noise. 

A spread in \ARIM\ curves indicates that the probabilistic distance of \RIM\ values w.r.t. the ideal for all controllers increases at different rates. So, the algorithm represented by the slowest growing curve is the best to find robust controllers.

Overall, SNOBFit's and Nelder-Mead's \ARIM\ curves at various training noises perform similarly to L-BFGS across all problems. However, there are distinctions in the region of $\sigma_\text{sim} \leq 0.05$ where L-BFGS curves start at lower \ARIM\ values and grow more quickly compared to SNOBFit curves at various noise levels. In the region of $\sigma_\text{sim} \geq 0.05$ the SNOBFit curves comparatively grow more slowly, possibly because the fidelity has degraded so much that further deterioration is less likely across all $100$ controllers. The Nelder-Mead curves exhibit similar behavior to the SNOBFit curves in that there is less variance w.r.t. the $\sigma_\text{train}$ levels, both when overall performance is good and when it is poor.

Compared to other algorithms, there is more variance in the PPO \ARIM\ curves across training noises for a particular spin transfer problem, with some curves overlapping each other. The best performing \ARIM\ curve is PPO at $\sigma_\text{train} = 0.05$ for the end-to-end transition shown in Fig.~\ref{fig:arimall}(f) (and for $6$ of $8$ cases in App.~\ref{sec:arim_allspins}). 
This indicates that PPO is often capable of finding robust solutions, but the optimal value of training noise varies across the transition problems.

We also present an extended \RIM\ analysis (like in Sec.~\ref{sec:individualcont}) for the controllers found for the same transition problem at training noises for the derivative-free approaches. The \RIM{}s at $\sigma_\text{sim} \in {0, \dotsc, 0.1}$ are plotted in Fig.~\ref{fig:arimall}(g)-(i) for PPO, SNOBFit and Nelder-Mead at $\sigma_\text{train} = 0.03$. On an individual level, SNOBFit and Nelder-Mead controllers share more algorithmic robustness and fidelity characteristics with each other across $\sigma_\text{train}$ than with PPO controllers: i.e., they have high \RIM\ variance within distribution per $\sigma_\text{train}$. This performance is also comparable to the L-BFGS controllers shown in Fig.~\ref{fig:noiselessindividualconts}(d). On the other hand, individually, the controllers found by PPO differ significantly across $\sigma_\text{train}$ where notably the \RIM\ and \ARIM\ values stay uniformly very low for the case $\sigma_\text{train} = 0, 0.03$ and the controllers are generally distinctly robust compared to SNOBFit and Nelder-Mead controllers.

Finally, we suggest possible explanations for these differences in behavior between algorithms. Since SNOBFit constructs local quadratic models to estimate gradients, it effectively filters out the perturbations $S_{\sigma_\text{train}}$. The manifestation of this effect is that the controllers at one training noise react similarly w.r.t. the \RIM, compared to controllers at other training noises (including the case of no training noise) as well as controllers found by L-BFGS. For Nelder-Mead, there are fewer noise-adaptation mechanisms compared to PPO and SNOBFit for large noise perturbations that might affect the quality of the estimated gradient direction and hence the rate of growth of the \ARIM\ w.r.t. simulation noise at higher training noise levels is unavoidable.

In contrast, PPO does not filter out the perturbations under $S_{\sigma_\text{train}}$ and forms its policy gradient estimates from stochastic fidelity function evaluations, which likely differentiates it from SNOBFit.  PPO also effectively estimates the fidelity landscape non-linearly using a fixed two-layer linear ($100 \times 100$ dimensional) neural-network, which may lead to generally better \ARIM performance.

\subsection{Comparison of Control Algorithms with Unconstrained Resources}\label{sec:uncons_algocomparison}
\begin{figure*}[t]
  \includegraphics[width=1\textwidth]{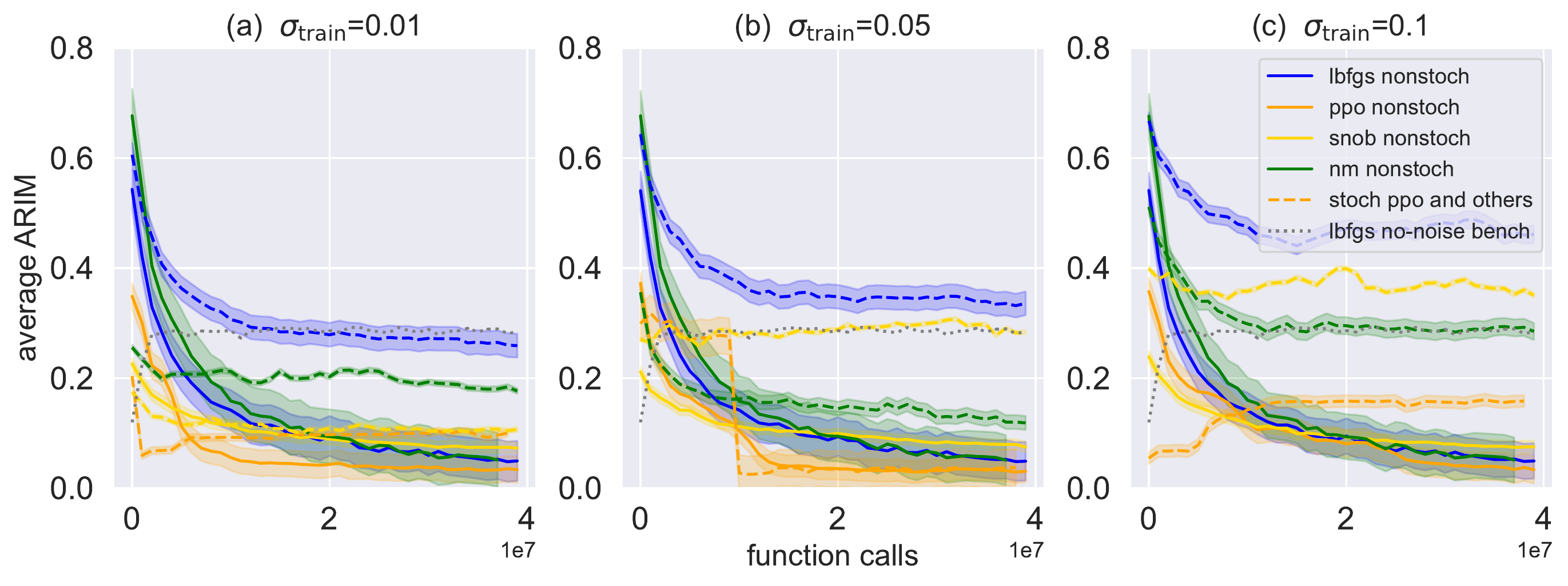}
  \caption{Asymptotic \ARIM\ performance when the nunber of objective function calls is unconstrained for the $M = 5$ end-to-middle spin transfer problem. The \ARIM{}s are averaged over the $\sigma_\text{sim}$ set $\{0, 0.01, \dots, 0.1\}$. The stochastic objective setting (i) is shown with dashed lines and deterministic \RIM\ objective setting (ii) with solid lines; different algorithms correspond to different colors. An L-BFGS no-noise benchmark is shown with a dotted line. The target \RIM\ is computed using $100$ non-stochastic fidelity evaluations. The \ARIM\ is computed and averaged over a $\sigma_\text{sim} = {0.0, 0.01, \dotsc, 0.1}$ set. Plots (a)-(c) correspond to training noise $\sigma_\text{train}=0, 0.05, 0.1$, where the curves are for $100$ controllers ranked by the corresponding objective function evaluation (i/ii) and are updated every $10^6$ function calls. For setting (ii), all control algorithms asymptotically reduce the average \ARIM, but this is not cost competitive with the stochastic setting (i) where PPO performance reaches the local minimum for all noise levels with fewer function calls. We see that the training noise level can help the landscape exploration process; this positively affects PPO in (a), (c) and Nelder-Mead in (b). For setting (i), L-BFGS, then SNOBFit, then Nelder-Mead, then PPO is the most prone to performance deterioration w.r.t. $\sigma_\text{train}$ due to the differences in their reliance on (estimated) gradient information. In these plots, the shading indicates 95\% confidence intervals, determined by using bootstrap resampling.}\label{fig:unconsopt}
\end{figure*}

We consider the behavior of the aforementioned control algorithms with an unconstrained number of objective function calls to address our motivating question (D), that is, we seek to understand an algorithm's ability to find robust controllers via the \ARIM\ -- without the function call constraint. Furthermore, we wish to ascertain what the effect of the training noise level $\sigma_\text{train}$ is on \ARIM\ optimization.

We consider two objective function settings: 
(i) \emph{stochastic objective}: for each evaluation, a new Hamiltonian is drawn according to the noise model, which corresponds to $1$ $S_{\sigma_\text{train}}$ perturbation in $\mathcal{F}$ during a single evaluation;
(ii) \emph{non-stochastic objective}: where the evaluation is over $k$ perturbed, but fixed, Hamiltonians, pre-drawn from the noise model such that optimization objective is a deterministic \RIM\ computed from $k$ fixed training perturbations $\{S^{(i)}_{\sigma_\text{train}}\}_1^{k}$. In this case, the function calls are counted as $k$ as they amount to $k$ different fidelity function evaluations per 1 optimization objective call.  Furthermore, since we cannot compute the analytical gradient of both objective functions, in order to use L-BFGS~\cite{lbfgs} in both settings, we use a version of L-BFGS that approximates the Hessian using forward differences.


We fix the control problem to be the end-to-middle $M=5$ transition. We consider the change in average \ARIM\ over $\sigma_{\rm sim} \in \{0,0.01,\dotsc, 0.1\}$ for the top $100$ controllers w.r.t. function calls. Three training noises $\sigma_\text{train} = 0, 0.05, 0.1$ are considered. The controller rankings are maintained w.r.t. the objective function and are updated in steps of $10^6$ function calls up to $4 \times 10^7$. For the stochastic setting (i), we maintain the controller ranking via the stochastic fidelity function evaluation. For the non-stochastic setting (ii) we maintain the ranking through the deterministic \RIM\ obtained using $k=100$ pre-drawn training perturbations $\{S^{(i)}_{\sigma_\text{train}}\}_1^{100}$. The choice of the hyperparameter $k$ was obtained using cross-validation. Specifically, for a particular training noise and control algorithm, we picked a $k$ from $\{10, 100, 10000\}$ to compute a \RIM\ in the objective function and then compared it to a \RIM\ computed using $k'=10000$ different training perturbations $\{S^{(i)}_{\sigma_\text{train}}\}_1^{10^4}$ that comprise a large validation set during the optimization run for all 100 controllers. We found no significant empirical difference in error between the objective function RIMs for $k=100$ and $k=10000$. Note that the variance in the RIM decreases as $O(1/k)$ by the law of large numbers.

For the non-stochastic setting (ii), it can be seen from Fig.~\ref{fig:unconsopt}(a)-(c) (solid lines) that the average \ARIM\ of all algorithms reduces asymptotically with the number of function calls. PPO attains the lowest final average \ARIM\ values at $4 \times 10^7$ function calls for each $\sigma_{\rm train}$ but its final \ARIM\ is not markedly better than the other algorithms considered here. 
However, this setting is quite expensive in terms of the total number of function calls.

For the stochastic setting (i) (dashed lines in Fig.~\ref{fig:unconsopt}(a)-(c)), increasing the training noise reduces the ability of the control algorithm to find robust controllers for all the $\sigma_\text{train}$, and the average \ARIM\ is not minimized to the same extent as in setting (ii). This makes sense, since the stochastic objective (i) is a noisy fidelity with a reduced focus on robustness. The average \ARIM\ is no longer reliably improved by any control algorithm in setting (i). 

Within setting (i), we note that PPO converges to the lowest average \ARIM\ value compared to the rest of the algorithms. This is theoretically justifiable as PPO optimizes a discounted \RIM\ by design (see App.~\ref{sec:ppooptrim}). Another thing to note is that lowest average \ARIM\ values obtained by PPO in all 3 settings are similar, even though they are not attained at the same number of function calls. This suggests that PPO's \ARIM\ performance can be made independent of the training noise levels, given an unconstrained number of objective function calls. However, this might be difficult to achieve completely since, even for PPO, there is a selection bias for low infidelity, but not low \RIM. This manifests itself in the fact that the average \ARIM\ starts increasing, albeit slowly, w.r.t. number of function calls after the lowest average \ARIM\ value is reached. Furthermore, we note that sharp transitions like the stochastic PPO curves depicted in Fig.~\ref{fig:unconsopt}(b) are also typically reported in classical reinforcement learning contexts and are linked to sharp improvements in the reward by the algorithm~\cite{bartosudden}.


To compare these results with the more standard noiseless fidelity maximization as a benchmark, we also plot the average \ARIM\ for L-BFGS with a noiseless fidelity objective function and analytical gradient information. This version of L-BFGS accumulates sharp peaks in the fidelity landscape with more function calls since it is gradient-based and is effectively climbing to the sharpest peak in the fidelity landscape. Hence its average \ARIM\ flatlines quickly w.r.t. function calls to a higher value compared to the other control algorithms in setting (i), with the exception of the forward differencing L-BFGS. 

Contrasting settings (i) and (ii) for a single control algorithm, the point at which there is an advantage for non-stochastic optimization via setting (ii) is around $10^7$ function calls for the algorithms, excluding L-BFGS with noise. For the regime below $10^7$ function calls, setting (i) has a clear advantage over (ii) for PPO and SNOBFit.

\section{Conclusion}

We have presented the robustness-infidelity measure ($\RIM_p$), a statistical generalization of the infidelity in the robustness sense, defined w.r.t. perturbations of arbitrary noise level $S_{\rm \sigma}$ in the fidelity function. The $\RIM_p$ is the $p$-th order Wasserstein distance of the infidelity distribution induced by $S_{\rm \sigma}$ from some ideal distribution that is impervious to $S_{\rm \sigma}$. We show that it is the $p$-th root of the $p$-th raw moment of the infidelity distribution and can be evaluated using perturbed fidelity function evaluations in physical experiments or Monte Carlo simulations. For $p=1$, the infidelity measure $\RIM_1$, reduces to the average infidelity. Using a metrization argument, we justify why $\RIM_1$ is a practical robustness measure for quantum control problems due to the convergence of $\RIM_p$ values, for all $p$, given highly robust and high fidelity controllers. The $\RIM_1$ also has a nice interpretation as the area under the curve of the cumulative distribution of the infidelity. We leave further analyses of the utility of the $\RIM_p$ for, e.g., the optimization of robustness, for future work.

The \RIM\ is further generalized to define an algorithmic \RIM\ (\ARIM) to compare the performance of control algorithms in terms of their ability to find robust high-fidelity controllers. Even though the \RIM\ and \ARIM\ are illustrated for static controls, they can be computed in any situation that generates a fidelity distribution over $[0,1]$, including time-dependent controls and open quantum systems, enabling their use and further study for a wide range of practical quantum control problems.

We have used the \RIM\ under model and controller noise to quantify the performance, in terms of the robustness and fidelity, of individual controllers for excitation transfer in spin chains by energy landscape shaping. The controllers were obtained by four control algorithms (PPO, SNOBFit, Nelder-Mead, L-BFGS) at simulation noise scales of up to $10\%$. Using the \RIM\, we found that high-fidelity controllers can vary widely in robustness to noise across all algorithms that we studied, although there are notable differences in algorithmic efficacy w.r.t. robustness, as indicated by the \ARIM. We also demonstrate a consistency statistic that can be used to differentiate control algorithms by how correlated their controller infidelities are with the ${\rm RIM}_1$. This provides a method to predict robustness via the ${\rm RIM}_1$ without its explicit evaluation.

To compare the control algorithms, we studied their \ARIM\ performance for multiple spin transfer problems. Under constrained function calls of a stochastic objective function (noisy infidelity), PPO performed better than SNOBFit and Nelder-Mead at certain, problem-specific training noise levels. SNOBFit performance at different training noise levels was similar, regardless of whether it was good or bad, suggesting that it is filtering out the noise.  Nelder-Mead exhibits similarly consistent behavior across training noise levels with less than optimal performance for all but one problem. With unconstrained stochastic function calls, PPO showed excellent performance compared to the other algorithms, independent of the training noise level, since its reward accumulation strategy implicitly optimizes a discounted \RIM. 

In contrast, when optimizing the $\RIM_1$ (average infidelity) over a fixed ensemble of perturbations, all algorithms were capable of asymptotically finding an optimum. However, this approach is expensive in terms of the number of function calls compared to the aforementioned stochastic optimization setting with a noisy fidelity function as the objective. Our results also show that for stochastic settings, e.g., shot noise, PPO (or more generally reinforcement learning) is a promising approach to obtain robust controllers.

A limitation of this work is that we require the computation of multiple controllers per control problem. In simulation, this further involves numerous time-consuming matrix exponential evaluations to generate a large number of samples per controller to approximate the \RIM\ measure. More work is necessary to elucidate the fundamental limitations of the optimization landscape. Nevertheless, our statistical robustness approach is a useful tool that can be applied in a wide range of quantum control scenarios where analytic approximations with small and/or uncorrelated noise are unsuitable.

\begin{acknowledgments}
I.\ Khalid acknowledges support for this work by a PhD scholarship from the School of Computer Science and Informatics, Cardiff University. The authors acknowledge the support of the Supercomputing Wales project to obtain the computational results presented herein, which is part-funded by the European Regional Development Fund (ERDF) via the Welsh Government.
\end{acknowledgments}

\bibliographystyle{apsrev4-2}
\bibliography{references}
\appendix
\section{\texorpdfstring{$\RIM_p$}{RIMp} Calculations}

\subsection{\texorpdfstring{$p$}{p}-th Order \RIM}\label{sec:rimcalculations}

{
In the subsequent argument, recall that the quantile function is the inverse of the CDF function. Following Thm.~\ref{th:quantileeqiv}, we can write the $\RIM_p$ as
\begin{equation}
  \RIM_p = \left(\int_0^1{|Q_{\mathbf{P}(\mathcal{F})}(z)-Q_{\delta_1}(z) |^p}\,dz\right)^{\frac{1}{p}}.
\end{equation}
Note that both terms in the integrand are $0$ at $z=0$. Then, for $z \in (0,1]$, by definition,
\begin{align}
  Q_{\delta_1}(z) 
    =& \inf\{x \in \mathbb{R}: C_{\delta_1}(x) \geq z > 0\} \nonumber\\
    =& \inf\{x \in \mathbb{R}: C(x=1) \geq z > 0\} \nonumber\\ 
     & \text{using the CDF of $\delta_1$ in Eq.~\eqref{eq:deltafunc_cdf}} \nonumber\\
    =& 1. 
\end{align} 
Another way to see this is to use the inverse property $Q_{\delta_1}(\cdot) = C_{\delta_1}^{-1}(\cdot)$. The CDF is $0$ in the interval $[-\infty, 1)$ and $1$ in $[1,\infty]$. Next, we perform a change of variable $z=C_{\mathbf{P}({\mathcal{F}})}(f)$. The differential is given by, $dz = \frac{dC_{\mathbf{P}({\mathcal{F}})}(f)}{df}df = \mathbf{P}(\mathcal{F}=f)df$ as the derivative of the CDF w.r.t. the random variable is the probability distribution function. Substituting the terms, we get
\begin{equation}
  \RIM_p = \left(\int_0^1{|Q_{\mathbf{P}({\mathcal{F}})}\left(C_{\mathbf{P}({\mathcal{F}})}(f)\right)-1 |^p}\mathbf{P}(\mathcal{F}=f)\,df\right)^{\frac{1}{p}}.
\end{equation}
Now we use the fact that $Q_{\mathbf{P}({\mathcal{F}})}(C_{\mathbf{P}({\mathcal{F}})}(f))=f$ (inverse property) to obtain, 
\begin{equation}
  \RIM_p = \left(\int_{0}^1{\mathbf{P}(\mathcal{F}=f)|f-1|^p}\,df\right)^{\frac{1}{p}}.
\end{equation}
Since the domain of integration remains invariant, for fidelity measures with support in $[0,1]$, it can be extended to $[-\infty, \infty]$. We obtain,
\begin{align}\label{eq:porderRIM}
  \RIM_p 
  =& \left(\int_{-\infty}^\infty{\mathbf{P}(\mathcal{F}=f)|f-1|^p}\,df\right)^{\frac{1}{p}}\nonumber\\
  =& \left(\int_{-\infty}^\infty{\mathbf{P}(\mathcal{F}=f)(1-f)^p}\,df\right)^{\frac{1}{p}}\nonumber\\
   & \text{as } f \leq 1\text{, switch the order and drop } |\cdot|\nonumber\\
  =& \mathds{E}_{f\sim \mathbf{P(\mathcal{F})}}\left[(1-f)^p\right].
\end{align}
We obtain the last line using the expectation operator $\mathds{E}_{f\sim \mathbf{P(\mathcal{F})}}[(\cdot)] := \int (\cdot)\mathbf{P}(\mathcal{F}=f)\,df$. For $p=1$, using Eq.~\eqref{eq:porderRIM}, $\RIM_1 = 1-\mathds{E}_{f \sim \mathbf{P(\mathcal{F})}}\left[f\right]$.

We can further decompose the $\RIM_p$ as a sum of expectations of various powers of the fidelity, 
\begin{align}\label{eq:porderRIMbinomial}
  \RIM_p &= \left(\sum_{k=0}^p \binom{p}{k}(-1)^k \int_{-\infty}^\infty{\mathbf{P}(\mathcal{F}=f) f^k}\,df\right)^{\frac{1}{p}}\nonumber\\
  &\qquad \text{using the binomial theorem}\nonumber\\
  &= \left(\sum_{k=0}^p \binom{p}{k}(-1)^k \mathds{E}_{f\sim \mathbf{P(\mathcal{F})}}\left[f^k\right]\right)^{\frac{1}{p}}.
\end{align}
For example, using Eq.~\eqref{eq:porderRIMbinomial} for $p=2$, we obtain
}
\begin{align}
  \RIM_2 
  =&  \sqrt{1-2\mathds{E}_{f\sim \mathbf{P(\mathcal{F})}}\left[f\right]+\mathop{Var}(f)+ \mathds{E}{^2}_{f\sim \mathbf{P(\mathcal{F})}}\left[f\right]}\nonumber\\
   & \text{using Eq.~\ref{eq:porderRIMbinomial}, and} \nonumber\\
   & \mathop{Var}(X) = \mathds{E}_{X\sim\mathbf{P}}\left[X^2\right] - \mathds{E}^2_{X\sim\mathbf{P}}\left[x\right]\nonumber\\
  =& \sqrt{\RIM_1+\mathop{Var}(f)-\mathds{E}_{f\sim \mathbf{P(\mathcal{F})}}\left[f\right]\RIM_1}\nonumber\\
  =& \sqrt{\mathop{Var}(f)+\RIM_1^2}\\
   & \text{expanding } \RIM_1 \text{ and simplifying}.\nonumber
\end{align}
Likewise, we get
\begin{multline}
  \RIM_3 = \left(\RIM_1^3+3\mathop{Var}(f) \right.\\
                 \left. + \mathds{E}{^3}_{f\sim \mathbf{P(\mathcal{F})}}\left[f\right]- \mathds{E}_{f\sim \mathbf{P(\mathcal{F})}}\left[f^3\right]\right)^{\frac{1}{3}}.
\end{multline}
The degree of distinguishability of the fidelity distribution from the ideal becomes better for higher $p$ at the cost of the outliers becoming more influential.

\subsection{Error Bound on the \texorpdfstring{$\RIM_p$}{RIMp} and \texorpdfstring{$\ARIM_p$}{ARIMp} Estimators}\label{sec:dkwrimerror}

Here we propose a probably approximately correct (PAC) alternative error bound for an estimation $\widehat{\RIM}_p$ of $\RIM_p$ in Eq.~\eqref{eq:porderRIM} based on an empirical estimate $\widehat{\mathbf{P}}(\mathcal{F})$ of its generating probability distribution $\mathbf{P}(\mathcal{F})$. With probability at least $1-\delta/2$,
\begin{align}
  &|\widehat{\RIM}_p - \RIM_p |\nonumber\\
  &= \left| \mathds{E}_{f\sim \mathbf{\widehat{P}(\mathcal{F})}}\left[(1-f)^p\right]^{\frac{1}{p}}
  - \mathds{E}_{f\sim \mathbf{P(\mathcal{F})}}\left[(1-f)^p\right]^{\frac{1}{p}}  \right|\nonumber\\
  &\leq | \mathds{E}_{f\sim \mathbf{\widehat{P}(\mathcal{F})}}\left[(1-f)^p\right]
  - \mathds{E}_{f\sim \mathbf{P(\mathcal{F})}}\left[(1-f)^p\right] |^{\frac{1}{p}}\nonumber\\
  &= \begin{multlined}[t]\left| \int_0^1{\widehat{\mathbf{P}}(\mathcal{F}=f)(1-f)^p}\,df \right.\hspace{11em}\\
      \left.-\int_0^1{\mathbf{P}(\mathcal{F}=f)(1-f)^p}\,df\right|^{\frac{1}{p}}\end{multlined}\nonumber\\
  &\leq \left( \int_0^1{|\widehat{\mathbf{P}}(\mathcal{F}=f)-\mathbf{P}(\mathcal{F}=f)|(1-f)^p}\,df\right)^{\frac{1}{p}}\nonumber\\
  &= \left( \int_0^1{\left| \widehat{\mathbf{P}}(\mathcal{F}=f)-\mathds{E}_{\widehat{\mathbf{P}} \sim \mathbf{D}}\left[\widehat{\mathbf{P}}(\mathcal{F}=f)\right]\right|(1-f)^p}\,df\right)^{\frac{1}{p}}\nonumber\\
  &\leq \frac{C^{\frac{1}{p}}}{p+1} = \frac{1}{p+1}\left(\frac{\log{\frac{4}{\delta}}}{2n}\right)^{\frac{1}{2p}}
\end{align}
where the second line and the fourth line come from using the reverse triangle inequality and in the fifth line we rewrite the true distribution $\mathbf{P}(\mathcal{F})$ as $\mathds{E}_{\widehat{\mathbf{P}} \sim \mathbf{D}}\left[\widehat{\mathbf{P}}(\mathcal{F})\right]$ which is true for any unbiased empirical estimator. We use McDiarmid's inequality to obtain the bounding constant $C$ using the fact that the probability distribution $\mathbf{D}$ generates a family of random variable empirical distributional estimators $\widehat{\mathbf{P}_j} = \tfrac{1}{n}\sum_{i=1}^n{\delta_{f_i}}$
where we have the differences occurring only on the $k$-th coordinate,
\begin{equation}
  \left|\widehat{\mathbf{P}}(f_1, \dotsc, f_{k}, \dotsc, f_n) - \widehat{\mathbf{P}}(f_1, \dotsc, f_{k'}, \dotsc, f_n) \right| \leq \frac{1}{n}
\end{equation}
where $n$ is the number of samples. A similar bound can also be derived for the $\widehat{\ARIM}$ estimator. This error bound is similar to the DKW (Dvoretzky-Kiefer-Wolfowitz) bound for the ECDF and would suffice in generating the 95\% confidence intervals for Fig.~\ref{fig:arimall} without the need to do bootstrap resampling.

\subsection{Relative Order of \texorpdfstring{$\RIM_p$}{RIMp}}\label{sec:equivalence}

Using Lyapunov's inequality, stating that $\mathds{E}[|X|^q]^{1/q} - \mathds{E}[|X|^{p}]^{1/p} \geq 0$ for $q \geq p>0$ for some $\mathds{E}[|X|^t] < \infty$, we show that
\begin{align}
  &\RIM_q -\RIM_p \nonumber\\
  &= \mathds{E}_{f\sim \mathbf{P}(\mathcal{F})}\left[(1-f)^q\right]^{\frac{1}{q}}\nonumber
    - \mathds{E}_{f\sim \mathbf{P(\mathcal{F})}}\left[(1-f)^p\right]^{\frac{1}{p}}\nonumber\\
  &= \mathds{E}_{f\sim \mathbf{P}(\mathcal{F})}\left[\left|(1-f)\right|^q\right]^{\frac{1}{q}}
    - \mathds{E}_{f\sim \mathbf{P(\mathcal{F})}}\left[\left|(1-f)\right|^p\right]^{\frac{1}{p}}\nonumber\\
  &\geq 0.
\end{align}
For any $q \geq p \geq s>0$, it follows that $\RIM_q \geq \RIM_p \geq \RIM_s$. The converse is true without the $p$-th roots. The linearity of expectations implies that $\mathds{E}_{f\sim \mathbf{P}(\mathcal{F})}\left[(1-f)^p - (1-f)^q\right] \geq 0 \iff 0 < p \leq q$.

We can also derive a lower bound on $\RIM_p$. For some $p' \geq p$, we have
\begin{align}\label{eq:rimupperbound}
  \RIM_{p'} \leq \RIM_{p}^{\frac{p}{p'}}
    &= \mathds{E}_{f\sim \mathbf{P}(\mathcal{F})}\left[(1-f)^p\right]^{\frac{1}{p'}}\nonumber\\
    &= \frac{\mathds{E}_{f\sim \mathbf{P}(\mathcal{F})}\left[(1-f)^p\right]^{\frac{1}{p}}}{\mathds{E}_{f\sim \mathbf{P}(\mathcal{F})}\left[(1-f)^p\right]^{\frac{1}{p}-\frac{1}{p'}}}\nonumber\\
    &\leq \frac{\RIM_p}{\mathds{E}_{f\sim \mathbf{P}(\mathcal{F})}\left[(1-f)\right]^{1-\frac{p}{p'}}}\nonumber\\
    &\leq \frac{\RIM_p}{(\min_f{(1-f)})^{1-\frac{p}{p'}}}
\end{align}
where the relation in the second last line is obtained by applying Jensen's inequality and the final line is obtained from the observation that $\min_f{(1-f)} < \mathds{E}[1-f]~\forall f$. Note that this result still depends on the data. Higher orders $p$ and $p'$ of the \RIM\ are related to each other in a concave sense and when $p, p' \to \infty$, the \RIM{}s become more equivalent. Conversely, near perfect fidelity, all the \RIM{}s are converging to $0$, but the presence of an outlier fidelity sample strongly governs how much discrepancy there still is between a higher-order \RIM\ and a lower order \RIM. This discrepancy is still concavely dependent on $p$ and $p'$.

\begin{figure*}[t]
  \includegraphics[width=0.9\textwidth]{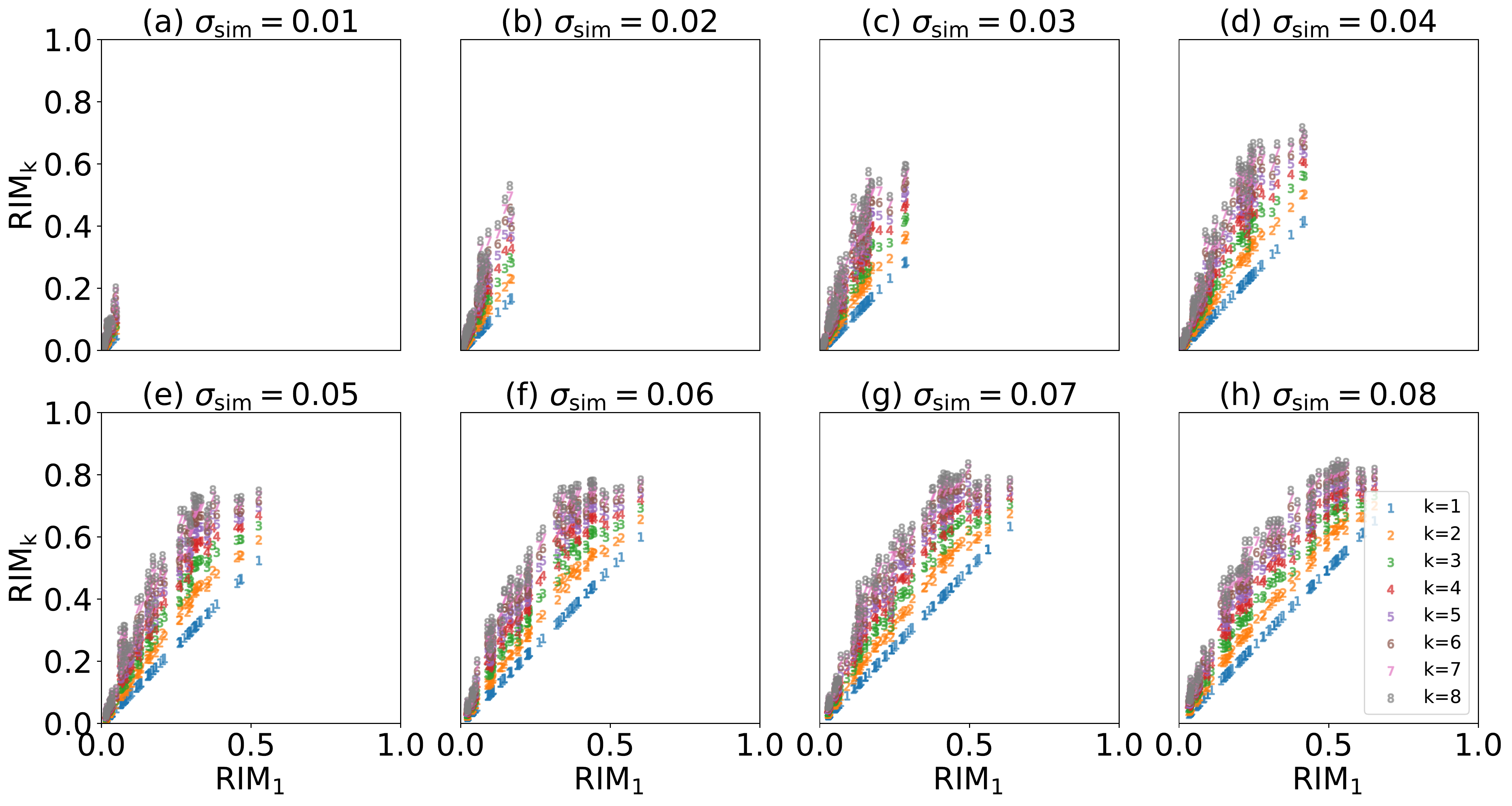}
  \caption{$\RIM_k$ scaling as a function of $\RIM_1$ for $100$ controllers for $M=5$ and the transition from $\ket{1}$ to $\ket{3}$ and $N=100$ samples per controller for \RIM\ evaluation. Each subplot (a)-(h) corresponds to a noise level $\sigma_\text{sim}$ indexing the fidelity probability distribution $\mathbf{P}_{\sigma_\text{sim}}(\mathcal{F})$. There is convergence and thus more agreement in $\RIM_k$ for small values.}\label{fig:rimk_scaling}
\end{figure*}

We arrive at equivalence relations for \RIM{}s of different order by noting that $\mathds{E}_{f\sim \mathbf{P}(\mathcal{F})}\left[(1-f)^p\right]
\geq m\sup_f(1-f) = m$ for the smallest positive finite measure $m>0$ on the domain set on which we define the probability distribution $\mathbf{P}(f)$. This follows from the continuity of $f$ and the continuity of $\mathbf{P}(f)$. If $f$ already has an ideal distribution, then this is trivially true.
Eq.~\eqref{eq:rimupperbound} yields
\begin{equation}
  \RIM_{p'} \leq {m}^{\left(\frac{1}{p'}-\frac{1}{p}\right)} \RIM_p.
\end{equation}
In practical settings, e.g., when using the ECDF, $m \geq \frac{1}{n}$. Intuitively, this follows from the observation that for any $\mathds{E}_{\mathbf{P}(X)}[X^p] = \int P(X) X^p \,dX  \approx \tfrac{1}{n} \sum_{i=1}^n X^p_i$ using samples $X_1, \dotsc, X_n$. For the estimated $\widehat{\RIM}_p$,
\begin{equation}
  \widehat{\RIM}_{p'} \leq {n}^{\left(\frac{1}{p}-\frac{1}{p'}\right)} \widehat{\RIM}_p.
\end{equation}
{
This implies that \RIM{}s of different orders are equivalent (in the convergence sense) when the bound holds.

Instead, motivated by the convergence in the Wasserstein distance, for fixed $n$, $\RIM_1$ can effectively constrain any $\RIM_p$ with $p>1$, since growth in $p$ is sublinear. This implies that the \RIM{}s converge when they tend to $0$ as seen in Fig.~\ref{fig:rimk_scaling}. We also note that the higher order \RIM{}s increase the measure's sensitivity to outliers greatly, even though growth in the \RIM\ is sublinear in $p$. For most practical purposes, the first order \RIM\ measure should be sufficient for performance measurements, especially in the quantum technologies setting.

\section{Optimization Algorithms}

\subsection{Implementation Details of the Optimization Objectives}\label{sec:optobjectives}

Details about the optimization objectives for the numerical results in Sec.~\ref{sec:numexp} are given in Table~\ref{table:imp_details}. In every section, for every $\sigma_\text{sim}$, the \RIM\ is evaluated using $N=100$ Monte Carlo $S_{\sigma_\text{sim}}$ perturbations to the fidelity function. The \RIM\ itself is only optimized in Sec.~\ref{sec:uncons_algocomparison} for the non-stochastic case (ii) where $100$ $S_{\sigma_\text{sim}}$ are sampled at the start and are reused for every function call.  Note, however, that we count these as $100$ function calls as these amount to $100$ fidelity function evaluations. Also, for better performance, in Sec.~\ref{sec:uncons_algocomparison} for the stochastic case (i), instead of using the analytical form for the gradient of fidelity $\mathcal{F}$, we use finite differences to approximate the gradients $\nabla_\Delta \mathcal{F}$ (where $\Delta$ are the controls).
}

\begin{table*}
  \caption{Implementation details for various optimization settings in the paper. For Sec.~\ref{sec:uncons_algocomparison}, the asymptotic setting, (i) refers to the stochastic scenario and (ii) refers to the non-stochastic scenario where the \RIM\ is optimized using the same $100$ fixed set of perturbations $\{S_{\sigma_\text{train}}\}$ per function call.}\label{table:imp_details}
  \begin{tabular}{|l|c|c|c|c|c|}
    \hline
    \textbf{Sec.}                      & \textbf{Obj.\ Function (OF) and args.}                  & \textbf{Train (OF) noise} & \textbf{Algorithm}  & \textbf{Total OF Calls} & \textbf{Single Call Cost}  \\ \hline
    \ref{sec:nonoise_contcomp}         & $\mathcal{F}$                                          &       No             & L-BFGS              & $10^6$   & $1$     \\ \hline
    \ref{sec:nonoise_contcomp}         & $\mathcal{F}$                                          &       No             & PPO                 & $10^6$   & $1$     \\ \hline
    \ref{sec:nonoise_contcomp}         & $\mathcal{F}$                                          &       No             & SNOBFit             & $10^6$   & $1$         \\ \hline
    \ref{sec:nonoise_contcomp}         & $\mathcal{F}$                                          &       No             & Nelder-Mead         & $10^6$   & $1$         \\ \hline
    \ref{sec:algocomparison}           & $\mathcal{F}$                                          &       No             & L-BFGS              & $10^6$   & $1$ \\ \hline
    \ref{sec:algocomparison}           & $\mathcal{F}$ \& 1 $S_{\sigma_\text{train}}$           &       Yes            & PPO                 & $10^6$   & $1$ \\ \hline
    \ref{sec:algocomparison}           & $\mathcal{F}$ \& 1 $S_{\sigma_\text{train}}$           &       Yes            & SNOBFit             & $10^6$   & $1$ \\ \hline
    \ref{sec:algocomparison}           & $\mathcal{F}$ \& 1 $S_{\sigma_\text{train}}$           &       Yes            & Nelder-Mead         & $10^6$   & $1$ \\ \hline
    \ref{sec:uncons_algocomparison}(i) & $\mathcal{F}$ \& 1 $S_{\sigma_\text{train}}$           &       Yes            & L-BFGS              & $\infty$ & $1$ \\ \hline
    \ref{sec:uncons_algocomparison}(i) & $\mathcal{F}$ \& 1 $S_{\sigma_\text{train}}$           &       Yes            & PPO                 & $\infty$ & $1$ \\ \hline
    \ref{sec:uncons_algocomparison}(i) & $\mathcal{F}$ \& 1 $S_{\sigma_\text{train}}$           &       Yes            & SNOBFit             & $\infty$ & $1$ \\ \hline
    \ref{sec:uncons_algocomparison}(i) & $\mathcal{F}$ \& 1 $S_{\sigma_\text{train}}$           &       Yes            & Nelder-Mead         & $\infty$ & $1$ \\ \hline
    \ref{sec:uncons_algocomparison}(ii)& \RIM\ \& 100 fixed $S_{\sigma_\text{train}}$           &       No             & L-BFGS              & $\infty$ & $100$ \\ \hline
    \ref{sec:uncons_algocomparison}(ii)& \RIM\ \& 100 fixed $S_{\sigma_\text{train}}$           &       No             & PPO                 & $\infty$ & $100$ \\ \hline
    \ref{sec:uncons_algocomparison}(ii)& \RIM\ \& 100 fixed $S_{\sigma_\text{train}}$           &       No             & SNOBFit             & $\infty$ & $100$ \\ \hline
    \ref{sec:uncons_algocomparison}(ii)& \RIM\ \& 100 fixed $S_{\sigma_\text{train}}$           &       No             & Nelder-Mead         & $\infty$ & $100$ \\ \hline
  \end{tabular}
\end{table*}

\subsection{PPO Optimizes a Discounted \texorpdfstring{$\RIM_1$}{RIM1}}\label{sec:ppooptrim}

We follow the standard finite-horizon Markov Decision Process (MDP) formulation for the reinforcement learning setting for states, actions and one-step state transition rewards $(s_t,a_t,r_t)$ that are sampled in trajectories $\tau = \{(s_t,a_t,r_t):t=1,\dotsc,T\}$ stored in the buffer $\mathbf{D}$. The proximal policy optimization (PPO) algorithm uses a clip objective to update the policy $\pi_\theta$ parameters $\theta$ with first-order constraints that minimize policy distributional divergence. The policy objective is
\begin{multline}\label{eq:policyopt}
  \theta_{k+1} \propto
    \argmax_{\theta} \sum_{\tau \in \mathbf{D}}\; \sum_{a_t, s_t, r_t \in \tau}^T\\
    \min \left[\frac{\pi_{\theta}(a_t|s_t)}{\pi_{\theta_k}(a_t|s_t)}A_{\pi_{\theta_k}}(s_t,a_t)\right.\\
    \left.\mathop{clip}(\epsilon, A_{\pi_{\theta_k}}(s_t,a_t))\right],
\end{multline}
where $\pi(\cdot)$ is the policy probability distribution. The advantage estimates are
\begin{equation}
  A_{\pi_{\theta_k}}(s_t,a_t) = \sum_{i=t}^{T-1}{(\gamma \lambda)^{i-t}(r_t+\gamma V_{\phi_k}(s_{t+1}) - V_{\phi_k}(s_t)) }
\end{equation}
with value function $V_\phi(s_t)=\mathds{E}_{\pi}\left[\sum_{i=0}^{T-1}\gamma^{i}r_{t+i+1}|s=s_t\right]$ where $\phi$ are the value function parameters. The value function is regressed onto discounted rewards sampled according to $\pi(\cdot)$. The clip function truncates the advantages to be between $(1\pm \epsilon)A_{\pi_{\theta_k}}$. The value function's optimization objective is
\begin{equation}
  \phi_{k+1} \propto \argmax_{\phi}\sum_{\tau \in \mathbf{D}} \sum_{t=0}^T \left(V_{\phi}(s_t)-\sum_{i=t}^{T}\gamma^{i}r_{i}(s_t^\tau)\right)^2.
\end{equation}
The algorithm tries to maximize this expression. In the case of flat rewards and advantages $\lambda=\gamma=1$, the advantage estimates are
\begin{multline}
  A_{\pi_{\theta_k}}(s_t,a_t) = \\
  V_{\phi_k}(s_t) - \left(V_{\phi_k}(s_T) + \sum_{i=t}^{T-1}{r_{t}}\right) =  V_{\phi_k}(s_t)-\widehat{V_{\phi_k}}(s_t).
\end{multline}
The value function can be written in terms of an expectation under the policy, as an average reward: $V_\phi(s_t)=T\mathds{E}_{\pi}\left[\frac{1}{T}\sum_{i=t}^{T}r_{i}|s=s_t\right]$. The optimal value function is defined by $V_*(s_t) = \max_{\pi}{V_{\phi}(s_t)}$, which is maximized if the policy is optimal, i.e., $\pi_\theta=\pi_{\theta^*}$ at $\theta=\theta^*$. Near optimality, the advantages are approximately $0$ as there should be no advantages conferred to the optimal policy $\pi_{\theta^*}$ which also has an optimal value function. Thus, $\widehat{V_{\phi^*}}(s_t) \to V_{\phi^*}(s_t)$ as $A_{\pi_{\theta^*}} \to 0$. The sample rewards minus the predicted rewards by the value function go to $0$ in Eq.~\eqref{eq:policyopt}. The same argument applies with discounts $\gamma, \lambda < 1$ and, hence, it can be shown that the algorithm optimizes a discounted $\RIM_1$ estimator as its value function. Most reinforcement learning algorithms effectively optimize the average or cumulative reward $\hat{J} \propto \sum_i r_i$ due to the one-step heuristic application of the Bellman principle of optimality~\cite{rl_no_effective_discounting}.

\subsection{More Consistency Statistic Plots}\label{sec:moreconst_stat_pppo}

\begin{figure*}
  \includegraphics[width=0.9\textwidth]{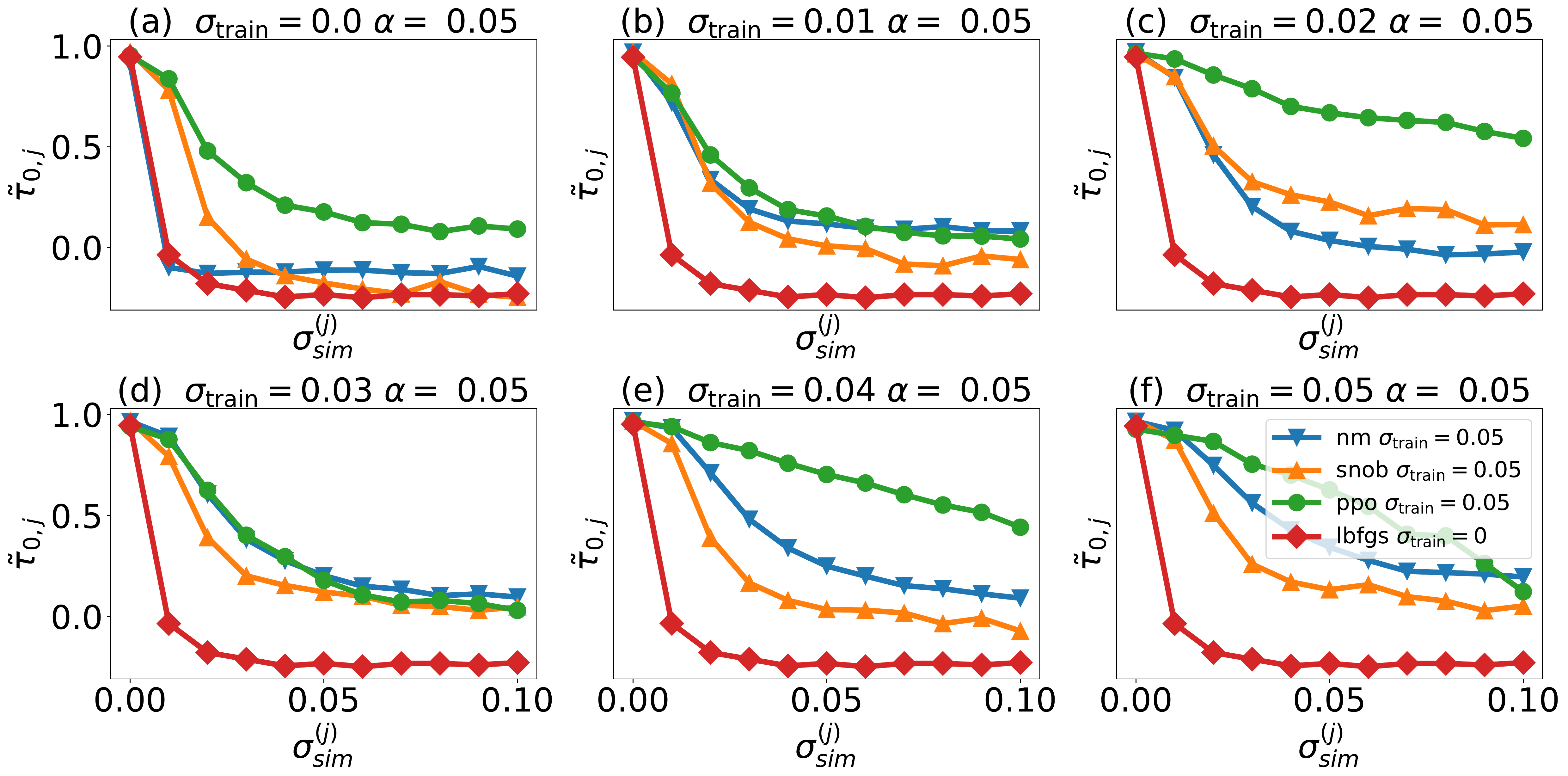}
  \caption{Consistency statistic $\tilde{\tau}_{0,j}$ for all algorithms at $\sigma_\text{train}=0.0, \dotsc, 0.05$ for discrepancy parameter $\alpha=0.05$ for $M=5$ and the transition from $\ket{1}$ to $\ket{3}$. Case (a) was presented in the main text. For (f), $\sigma_\text{train}=0.04$, PPO is actually more robust in terms of \ARIM\ growth compared with (e) as seen from their positions in Fig.~\ref{fig:arimall}(b). Characteristically, most low \ARIM PPO controllers 
  show high rank consistency in the region $0 \leq \sigma_\text{sim} \leq 0.04$. Nelder-Mead is similar to L-BFGS in all plots except (e) and (f), where it shows slightly more consistency than PPO and SNOBFit controllers.}\label{fig:ppo5to2tau}
\end{figure*}

\begin{figure*}
  \includegraphics[width=0.9\textwidth]{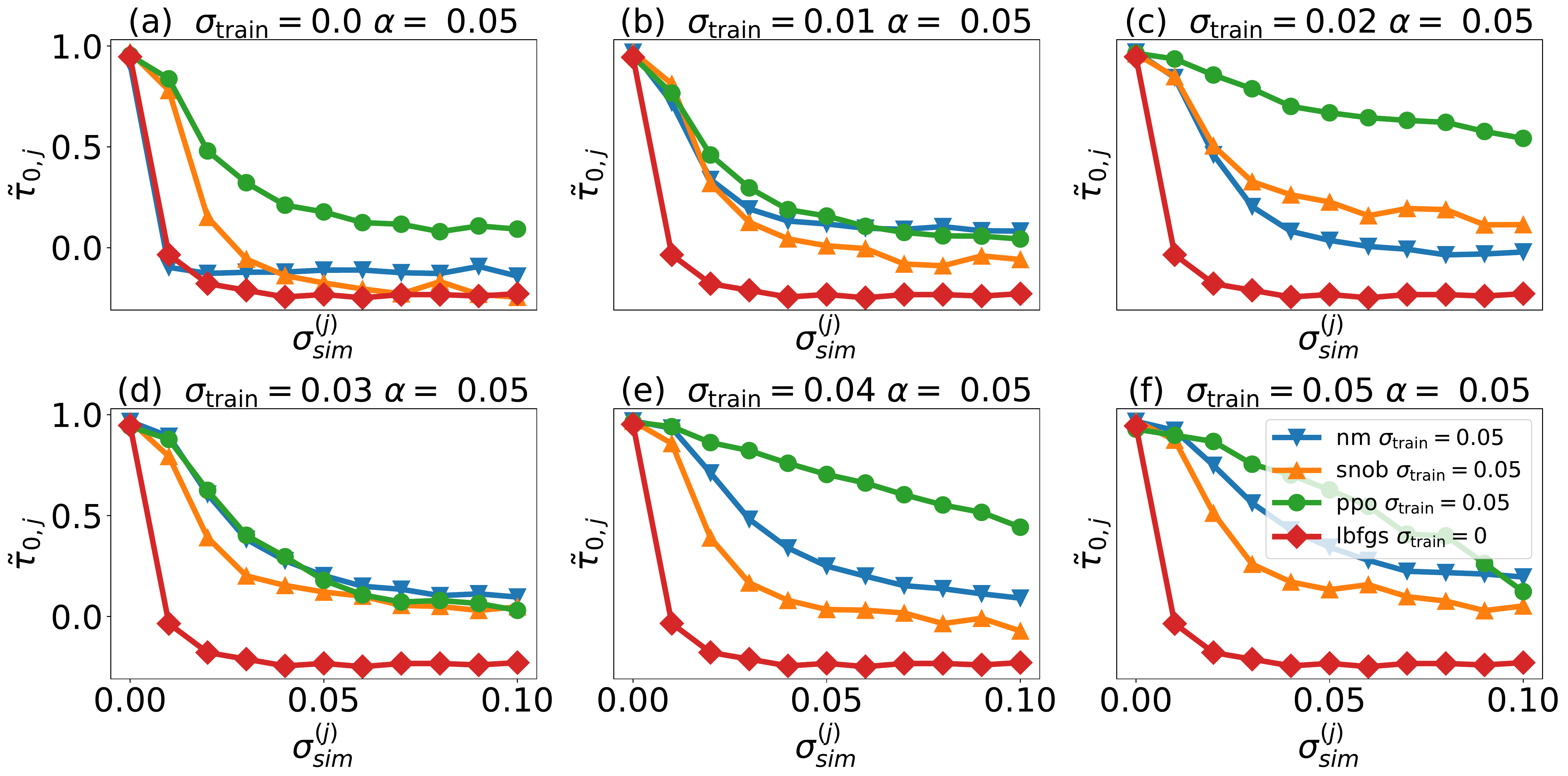}
  \caption{Consistency statistic $\tilde{\tau}_{0,j}$ for all algorithms at $\sigma_\text{train}=0.0, \dotsc, 0.05$ for discrepancy parameter $\alpha=0.05$ for $M=5$ and the transition from $\ket{1}$ to $\ket{4}$. Again, the PPO curves are the most consistent.}\label{fig:ppo5to4tau}
\end{figure*}

This section expands the discussion on the consistency statistic in the main text in Sec.~\ref{sec:consistencystat}. We plot the consistency statistic $\tilde{\tau}_{0,j}$ for all algorithms for $\alpha=0.05$ for the case $M=5$ and the transition $\ket{1}$ to $\ket{3}$ in Fig.~\ref{fig:ppo5to2tau}(a)-(f) ((a) is Fig.~\ref{fig:taumatrix}) and $\ket{1}$ to $\ket{4}$ in Fig.~\ref{fig:ppo5to4tau}(a)-(f) for multiple training noise levels. Note that for each subplot the L-BFGS curve is always the same at $\sigma_{\text{train}}=0$. The controllers found by PPO at $\sigma_\text{train} = 0.05$ are less consistent for some noise levels than others, e.g., $\sigma_\text{sim} \geq 0.04$ compared with the controllers found at $\sigma_\text{train} = 0.04$. This is also true for SNOBFit and Nelder-Mead. Moreover, the decline in the correlation values is smoothest for PPO compared to the rest for nearly all twelve instances shown in both figures. With more training noise, Nelder-Mead is sometimes closer in consistency to the controllers found to L-BFGS, e.g., Fig.~\ref{fig:ppo5to2tau}(a,b). But it produces more consistent controllers with increasing training noise likely due to diminishing returns of the gradient direction, makes its behavior more like SNOBFit and PPO.

For most PPO runs, the consistency statistic is highest for $\sigma_\text{sim} \leq 0.04$ and thus the infidelity rank order is a good predictor of \RIM\ rank order for higher $\sigma_\text{sim}$, which was not observed for any of the other algorithms. Also note that this analysis does not reveal anything about how high the \RIM\ values are for the controllers (a drawback of the non-parametric test) and should be processed as companion plots to the figures where these explicit values are shown.

\subsection{More Individual Controller Plots}\label{sec:extraplotscontcomp}

\begin{figure*}
  \includegraphics[width=0.64\textwidth]{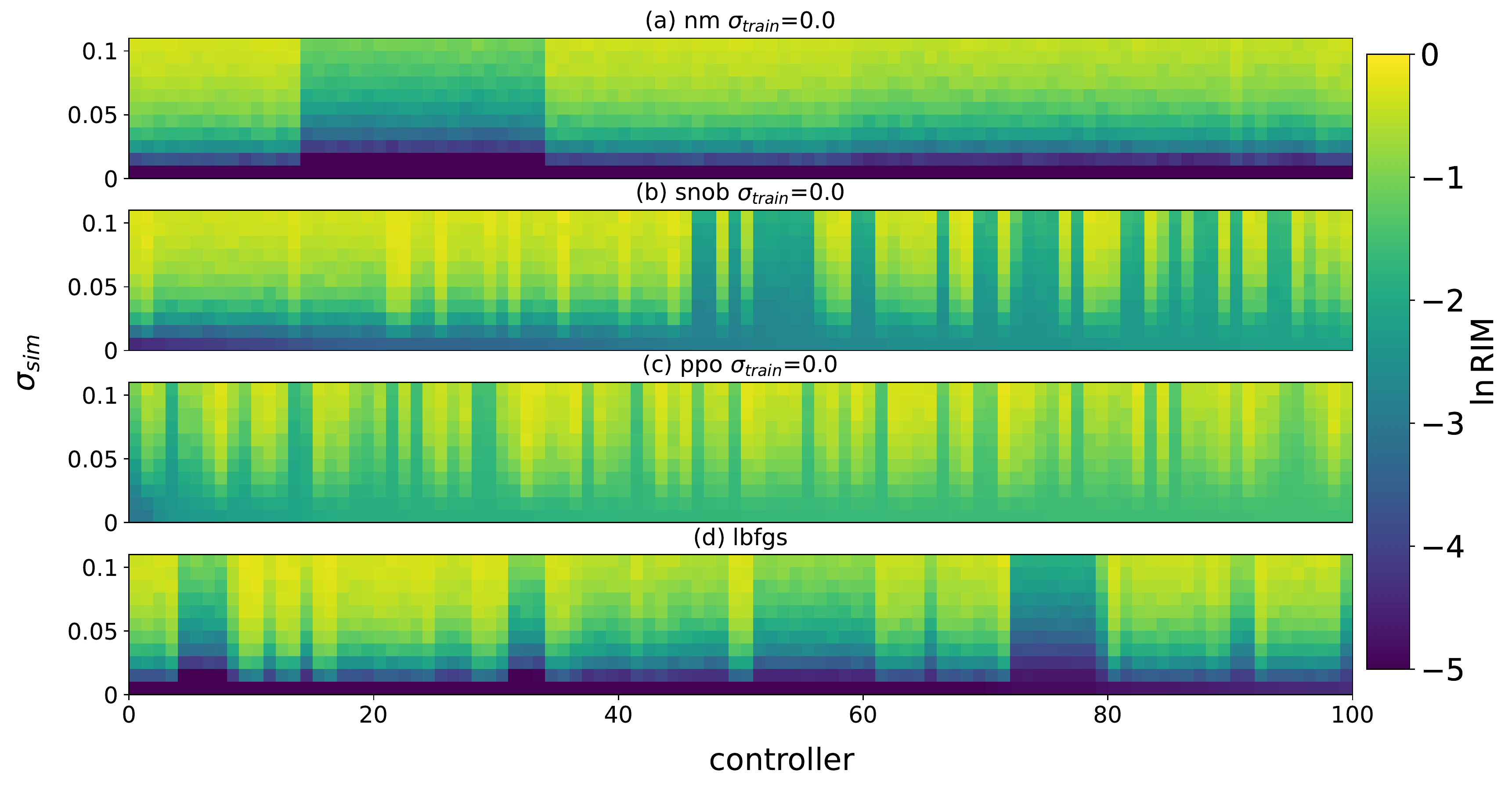}
  \hfill
  \includegraphics[width=.35\textwidth]{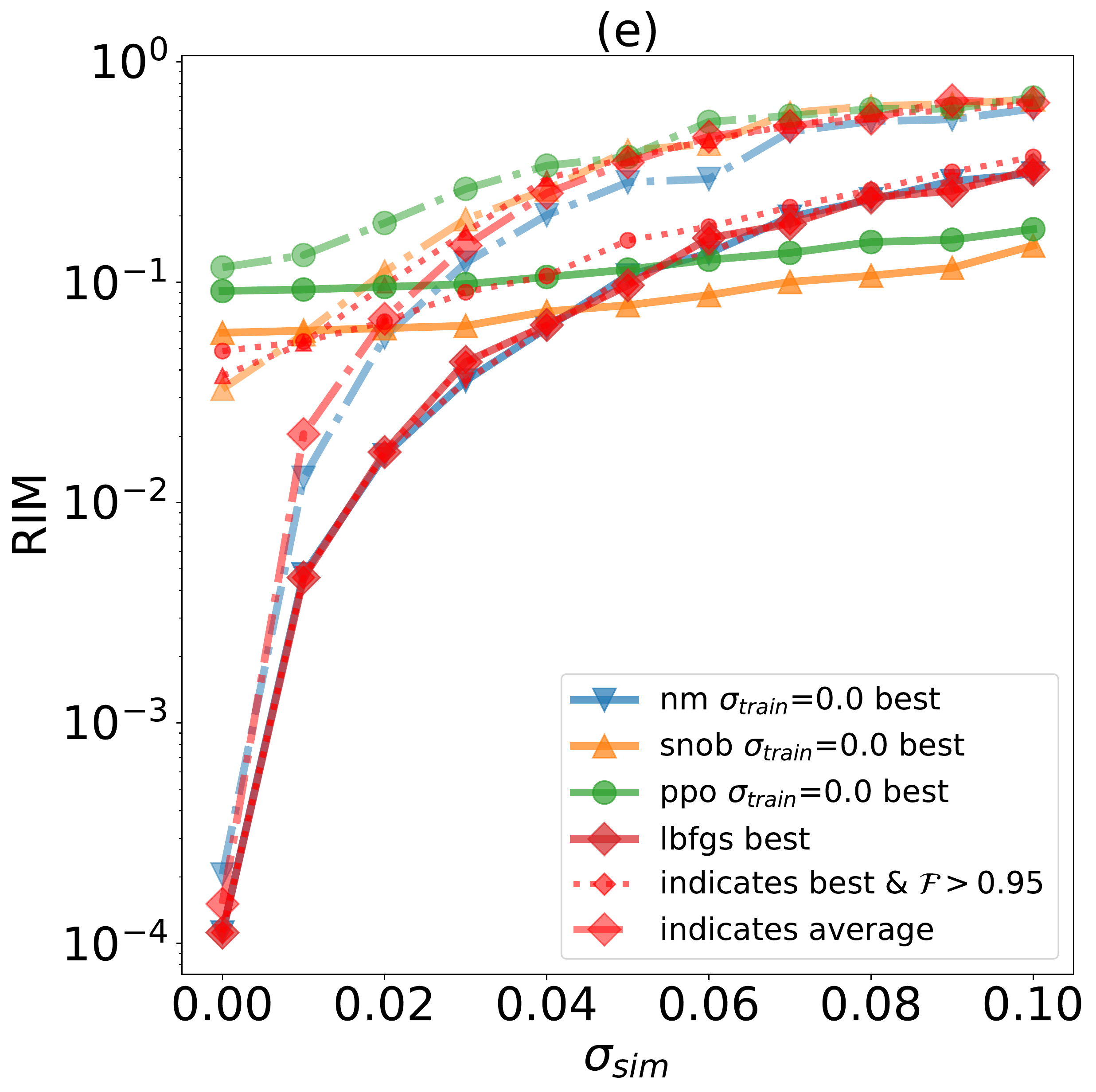}\\
  \caption{(a)-(d) $100$ controllers found for the XX spin chain model, Eq.~\eqref{eq:firstexcitationsubspaceXX}, using Nelder-Mead, SNOBFit, PPO ($\sigma_\text{train}=0$), and L-BFGS for $M=5$ and the spin transition from $\ket{1}$ to $\ket{5}$. All algorithms find controllers that are not very robust as indicated by the \RIM. PPO has notably worse initial infidelities for all controller compared to Fig.~\ref{fig:noiselessindividualconts}(c), but their degradation is slow as seen from (e). This is only the case for this noise level 
  and Fig.~\ref{fig:noisycontssupp}(r) indicates the existence of a much better controller set at $\sigma_\text{sim} = 0.05$ that is similar in performance to Fig.~\ref{fig:noiselessindividualconts}(c). From (e), we can see that Nelder-Mead and L-BFGS optimize the infidelity to $<10^{-4}$. However, these best controllers decay in robustness very quickly as well.
  }\label{fig:noiselessindividualcontssupp}
\end{figure*}

\begin{figure*}
  \includegraphics[width=1\textwidth]{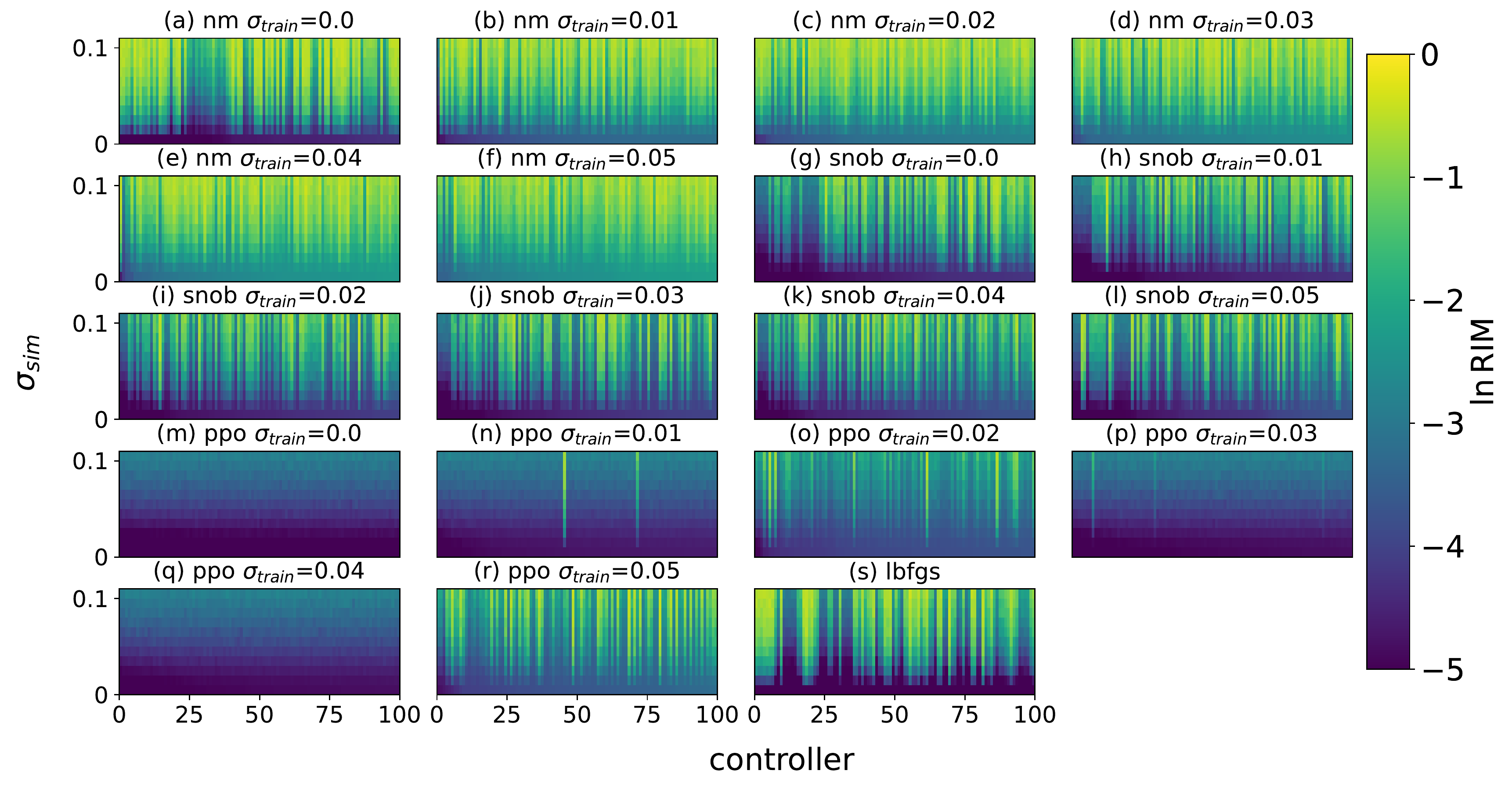}
  \caption{Individual-controller comparison between (a)-(f) Nelder-Mead, (g)-(k) SNOBFit and (m)-(r) PPO with $\sigma_\text{train}=0, 0.01, \dotsc, 0.05$, using $100$ controllers ranked by lowest infidelity (left) 
  for the case $M=5$ and the spin transition from $\ket{1}$ to $\ket{3}$. (s) shows the L-BFGS results for the same spin transition problem.}\label{fig:noisyconts}
\end{figure*}

\begin{figure*}
  \includegraphics[width=1\textwidth]{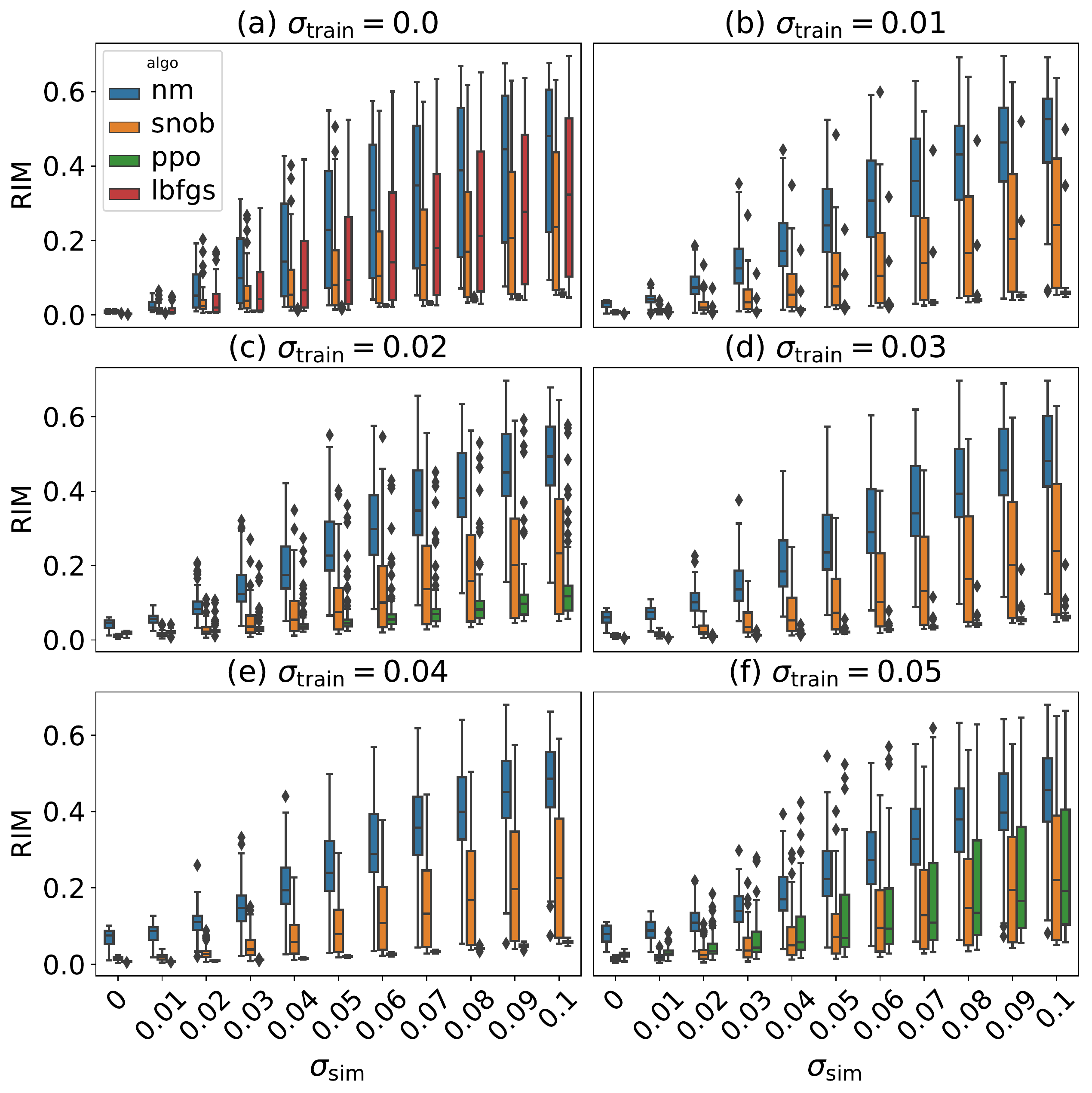}
  \caption{Box plots of the \RIM\ for the $100$ controllers for $M=5, \ket{1}$ to $\ket{3}$ shown in Fig.~\ref{fig:noisyconts} in the main text found by Nelder-Mead, SNOBFit, and PPO for various $\sigma_\text{train}$ (a)-(f). For the case $\sigma_\text{train}=0$ in (a), we also show L-BFGS box plots as a reference. On the distributional level, PPO controllers are generally the more robust of the three w.r.t. the \RIM, but there is high variance across $\sigma_\text{train}$ compared to the SNOBFit and Nelder-Mead controllers. The median SNOBFit \RIM\ value per $\sigma_\text{sim}$ is higher than L-BFGS, so it has a longer left tail.
  The Nelder-Mead controllers have the most weight on their right tails and are comparatively the worst.}\label{fig:boxcomp}
\end{figure*}

The results presented in Fig.~\ref{fig:noiselessindividualconts} ($M=5$ and transition $\ket{1}$ to $\ket{3}$) are not reflective of PPO's general behavior on the extended sample of problems examined in Sec.~\ref{sec:algocomparison}. Fig.~\ref{fig:noiselessindividualcontssupp} shows the case ($M=5$ and transition $\ket{1}$ to $\ket{4}$) where all the controllers found are not very robust. This is likely either due to unlucky sampling of the space of possible controllers or their non-existence. Note that SNOBFit and PPO are similar in their \RIM\ degradation as observed from Fig.~\ref{fig:noiselessindividualcontssupp}(e). We also provide some more cases ($M=5$ and transition $\ket{1}$ to $\ket{5}$) and ($M=6$ and transitions: $\ket{1}$ to $\ket{4}$, $\ket{1}$ to $\ket{6}$) for algorithm comparison of controllers under noisy training in Sec.~\ref{sec:algocomparison} to highlight some of the variation of controller quality for different regimes of noise and spin chain transitions observed in the main \ARIM\ comparison presented in Fig.~\ref{fig:arim_allspins}. Each individual subplot is the result of an independent run of each algorithm with a stochastic fidelity function evaluated under the unstructured perturbations using the same approach as earlier with $\sigma_\text{sim}$. These are also plotted for a more distributional comparison as pairwise box-plots in Fig.~\ref{fig:boxcomp}. For both, Figs.~\ref{fig:noisyconts} and~\ref{fig:boxcomp}, we also show L-BFGS results for comparison.

\subsection{Full ARIM comparisons}\label{sec:arim_allspins}

\begin{figure*}
  \centerline{\includegraphics[width=.75\textheight,height=\textwidth]{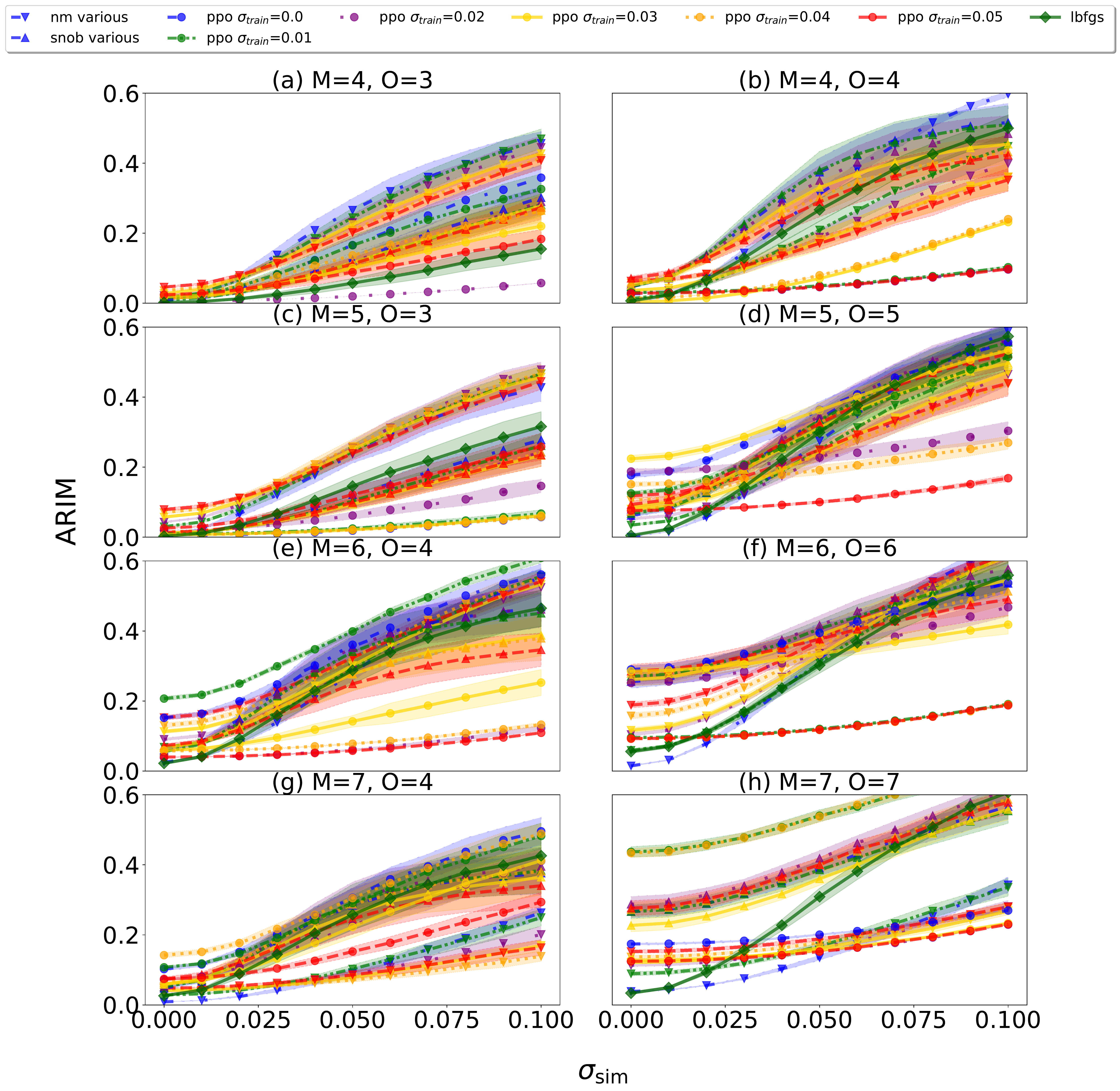}}
  \caption{\ARIM\ as a function of $\sigma_\text{sim}$ for $M=4,5,6,7$ where the left column contains end-to-middle transitions and the right column contains end-to-end transitions. The final state is denoted by $O$. The \ARIM\ is computed from a distribution of \RIM\ values for $100$ controllers for each $\sigma_\text{sim}$ for SNOBFit, Nelder-Mead, PPO and L-BFGS indicated by their marker shapes and line-styles. Both PPO and SNOBFit are run multiple times at $\sigma_\text{train}=0, 0.01,\dotsc, 0.05$ which is indicated by the color of the \ARIM\ curve. For all problems, PPO has higher variance with respect to $\sigma_\text{train}$ than SNOBFit and Nelder-Mead. The latter pair's performance curves are more in line with the L-BFGS curve for $\sigma_\text{sim} \geq 0.05$ and mostly worse for $\sigma_\text{sim} \leq 0.05$. For most of the problems the best performing (lowest) curve across all problems is PPO at $\sigma_\text{train} = 0.05$ (brown) except in (a) where it is PPO at $\sigma_\text{train} = 0.02$ and in (g) where it is Nelder-Mead at $\sigma_\text{train} \geq 0.04$. 95\% confidence intervals (shading) are computed using non-parametric bootstrap resampling~\cite{bootstrapresampling} with $100$ resamples.
  }\label{fig:arim_allspins}
\end{figure*}

{
For the cases $M=6,7$, both types of transitions appear to be challenging for PPO, SNOBFit and Nelder-Mead at most, if not all, training perturbation strengths; especially the end-to-end $M=7$ transition (Fig.~\ref{fig:arim_allspins}(d)), where PPO at $\sigma_\text{train} = 0.05$ is only marginally better than the rest of the algorithm runs, excluding Nelder-Mead. A pertinent question is whether this is genuinely reflective of the landscape or if, for PPO, our budget constraint of $10^6$ target functional calls is insufficient for larger system sizes, as the control problem is exponentially dependent on the number of control degrees of freedom. The former hypothesis might hint at a fundamental limitation on robustness of this particular control landscape. The fact that most noisy Nelder-Mead curves for these problems are clustering together suggests that noise could also help in reaching robust areas in the control landscape faster by regularizing or smoothing the landscape by an appropriate degree. We investigate asymptotic algorithm behavior w.r.t. the training noise in Sec.~\ref{sec:uncons_algocomparison} to illustrate this and show that there is convergence in PPO performance for all the noise levels at sufficiently many function calls.
}

\begin{figure*}
  \includegraphics[width=1\textwidth]{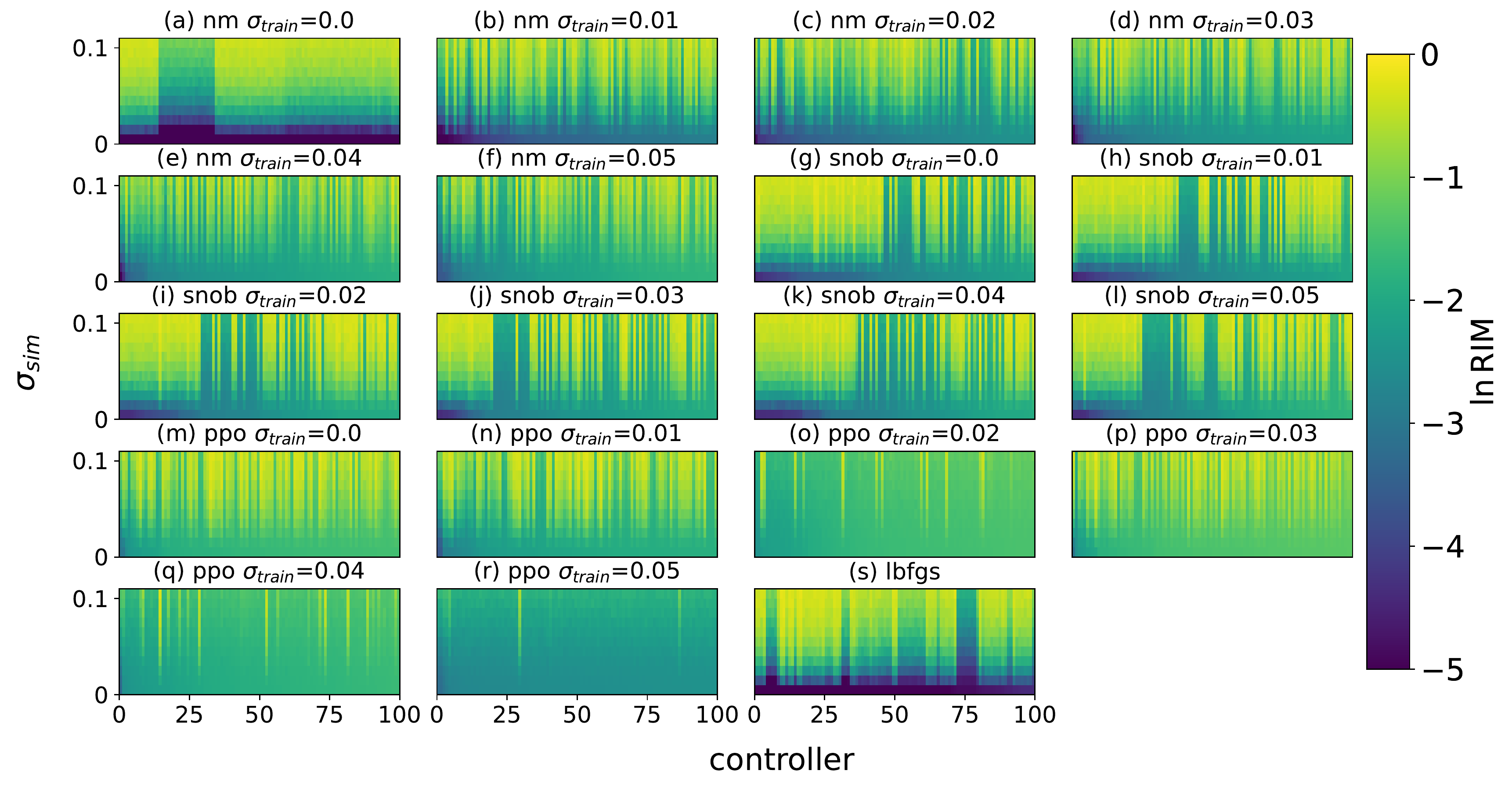}
  \caption{Individual-controller comparison between (a)-(f) Nelder-Mead, (g)-(k) SNOBFit, (m)-(r) PPO with $\sigma_\text{train}=0, 0.01, \dotsc, 0.05$, using $100$ controllers ranked by lowest infidelity for the case $M=5$ and the spin transition from $\ket{1}$ to $\ket{5}$. (s) shows the L-BFGS result for $\sigma_\text{train}=0$.}\label{fig:noisycontssupp}
\end{figure*}

\begin{figure*}
  \includegraphics[width=1\textwidth]{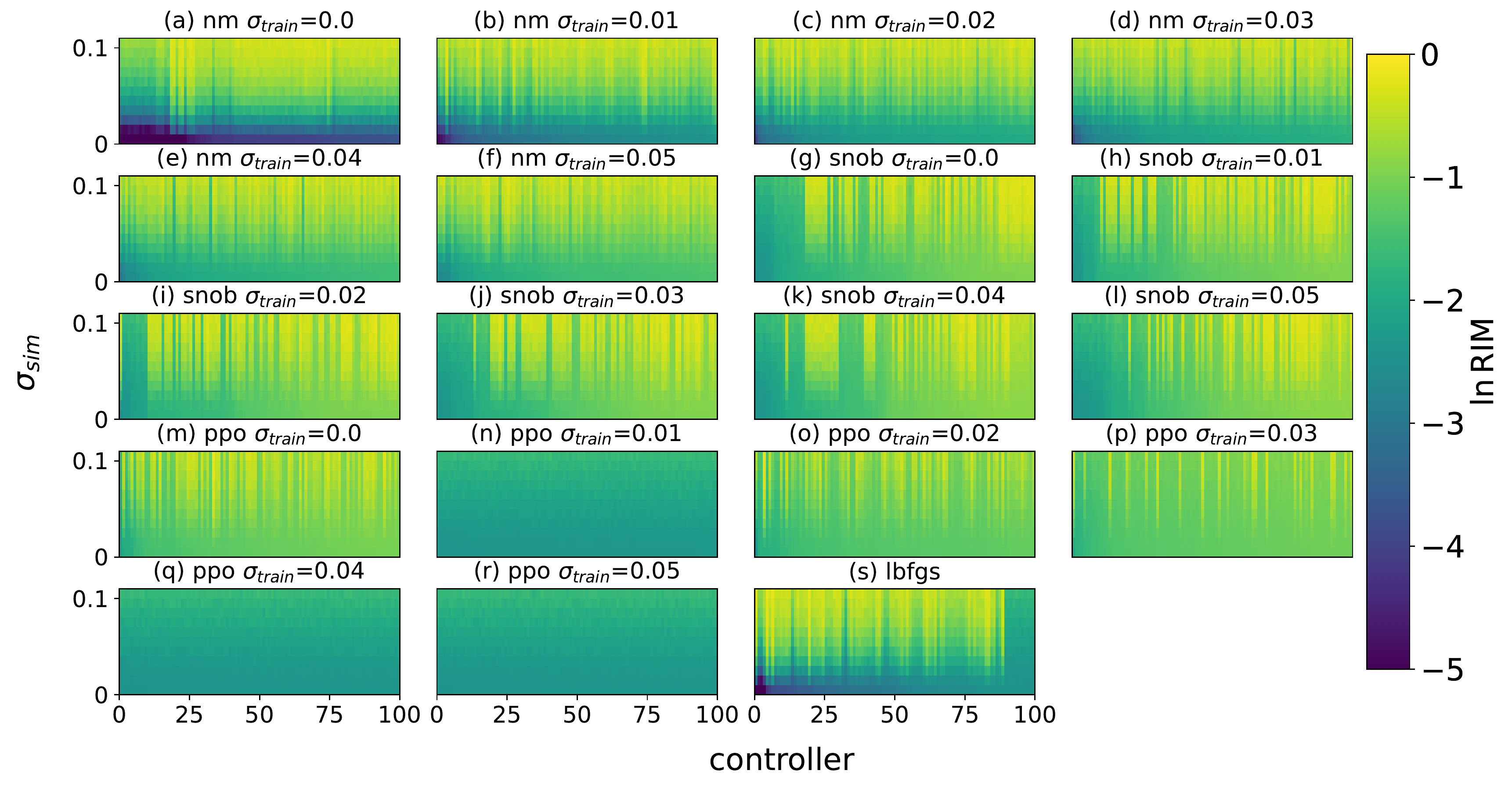}
  \caption{Individual-controller comparison between (a)-(f) Nelder-Mead, (g)-(k) SNOBFit, (m)-(r) PPO with $\sigma_\text{train}=0, 0.01, \dotsc, 0.05$, using $100$ controllers ranked by lowest infidelity
  for the case $M=6$ and the spin transition from $\ket{1}$ to $\ket{6}$. (s) shows the L-BFGS result for $\sigma_\text{train}=0$.}\label{fig:noisycontssupp2}
\end{figure*}

\begin{figure*}
  \includegraphics[width=1\textwidth]{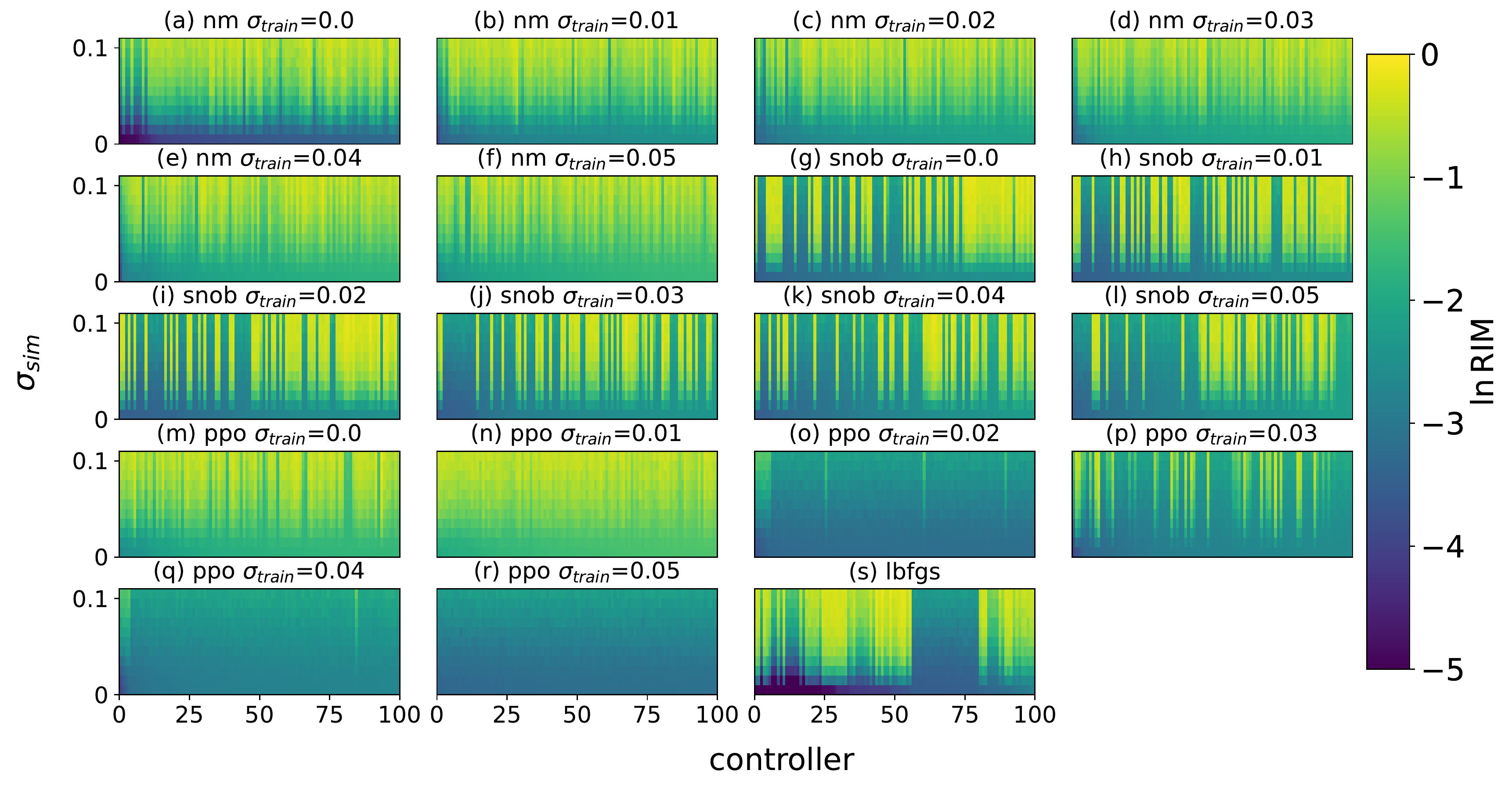}
  \caption{Individual-controller comparison between (a)-(f) Nelder-Mead, (g)-(k) SNOBFit, (m)-(r) PPO with $\sigma_\text{train}=0, 0.01, \dotsc, 0.05$, using $100$ controllers ranked by lowest infidelity
  for the case $M=6$ and the spin transition from $\ket{1}$ to $\ket{4}$. (s) shows the L-BFGS result for $\sigma_\text{train}=0$.}\label{fig:noisycontssupp3}
\end{figure*}

\end{document}